# Surrogate-Based Bayesian Inverse Modeling of the Hydrological System: An Adaptive Approach Considering Surrogate Approximation Error

**Jiangjiang Zhang**[1, 2], **Qiang Zheng**[1, 2], **Dingjiang Chen**[1, 3], **Laosheng Wu**[4], **and Lingzao Zeng**[1, 2]

[1] Institute of Soil and Water Resources and Environmental Science, College of Environmental and Resource Sciences, Zhejiang University, Hangzhou, China,

[2] Zhejiang Provincial Key Laboratory of Agricultural Resources and Environment, Zhejiang University, Hangzhou, China,

[3] Ministry of Education Key Laboratory of Environment Remediation and Ecological Health, Zhejiang University, Hangzhou, China,

[4] Department of Environmental Sciences, University of California, Riverside, CA, USA.

Corresponding Author: L. Zeng (lingzao@zju.edu.cn)

**Key Points:**

- Surrogate approximation error can introduce bias to surrogate-based inverse modeling
- We propose two strategies to quantify the surrogate approximation error and improve simulation accuracy
- The surrogate approximation error is gradually reduced to a small level in the high posterior density region

**Abstract**

Bayesian inverse modeling is important for a better understanding of hydrological processes. However, this approach can be computationally demanding as it usually requires a large number of model evaluations. To address this issue, one can take advantage of surrogate modeling techniques. Nevertheless, when approximation error of the surrogate model is neglected, the inversion result will be biased. In this paper, we develop a surrogate-based Bayesian inversion framework that explicitly quantifies and gradually reduces the approximation error of the surrogate. Specifically, two strategies are proposed to quantify the surrogate error. The first strategy works by quantifying the surrogate prediction uncertainty with a Bayesian method, while the second strategy uses another surrogate to simulate and correct the approximation error of the primary surrogate. By adaptively refining the surrogate over the posterior distribution, we can gradually reduce the surrogate approximation error to a small level. Demonstrated with three case studies involving high-dimensionality, multi-modality and a real-world application, it is found that both strategies can reduce the bias introduced by surrogate approximation error, while the second strategy that integrates two methods (i.e., polynomial chaos expansion and Gaussian process in this work) that complement each other shows the best performance.

# 1. Introduction

Uncertainty is ubiquitous in measuring, analyzing and predicting the system of concern in many different fields (Smith, 2014). In hydrological science, there is a growing interest in quantifying different sources of uncertainties with numerical modeling techniques, e.g., sensitivity analysis and data assimilation (Liu & Gupta, 2007; Song et al., 2015; Tartakovsky, 2013; Vrugt et al., 2008). To obtain an improved understanding of hydrological processes, it is necessary to reduce the uncertainty of model parameters by conditioning on measurements of model states with inverse modeling methods (Gupta & Nearing, 2014; Osorio-Murillo et al., 2015; Vrugt, 2016). Over the past decades, numerous inverse methods have been developed and/or applied in hydrological science, e.g., local and global optimization methods (Ahmad et al., 2014; Duan et al., 1992; Reed et al., 2013), ensemble Kalman filter and its variants (Chen & Zhang, 2006; Emerick & Reynolds, 2013; Evensen, 2009; Song et al., 2014; Zhang et al., 2018a), particle filter (Moradkhani et al., 2005; Vrugt et al., 2013; Weerts & El Serafy, 2006), and Markov chain Monte Carlo (MCMC) (Smith & Marshall, 2008; Vrugt et al., 2003; Vrugt et al., 2008; Zhang et al., 2015), just name a few.

In this paper, our concern is MCMC, which is a Bayesian inverse method that can sample the posterior distribution of model parameters in a consistent and coherent manner (Brooks et al., 2011; Vrugt, 2016). Over the past decade, MCMC has been widely used in hydrological simulations (Cao et al., 2018; Gaume et al., 2010; Goharian et al., 2018; Kuczera et al., 2010; Vrugt & Beven, 2018). However, when implementing MCMC for inverse modeling, a large number of model evaluations are usually required, especially in high-dimensional, strongly nonlinear problems. Here, we call the kind of model that is accurate enough for the task at hand the high-fidelity model (Peherstorfer et al., 2018). In many situations, the high-fidelity model is CPU-intensive, then the computational cost of MCMC simulation will be prohibitive.

To address the computational issue, one can take advantage of surrogate modeling techniques. A surrogate model (also called low-fidelity model, meta-model, proxy model, reduced model, or response surface, etc.) is a computationally cheaper model designed to approximate the responses of the high-fidelity model (Asher et al., 2015; Razavi et al., 2012). A surrogate model can be a data-driven model based on regression or interpolation, a reduced-order model obtained by projecting the original parameters and states into their lower-dimensional subspace, or a numerical model with a reduced precision, etc. (Smith, 2014). Compared to the high-fidelity model, the surrogate model can obtain similar (while less accurate) model outputs at a much lower computational cost. Due to their efficiency, different forms of surrogate models have been used in MCMC simulations of the hydrological system. For example, Zeng et al. (2012) used adaptive sparse grid interpolation, and Laloy et al. (2013) used generalized polynomial chaos expansion (PCE) to construct surrogates to accelerate posterior exploration of groundwater models. Inevitably, due to the existence of surrogate approximation error (i.e., the difference between the high-fidelity and the surrogate model outputs), using surrogate models directly in MCMC simulations will introduce some bias. To address this issue, Zeng et al. (2012) and Laloy et al. (2013) implemented two-stage MCMC simulations (i.e., sufficiently explored the parameter space with the surrogate model at stage one, then evaluated the high-fidelity model for correction at stage two), instead of only implementing the surrogate-based MCMC simulations.

Another strategy to handle the surrogate approximation error is to build a locally accurate surrogate in the posterior region. Nevertheless, this task is nontrivial since the position of true posterior region is unknown and to be estimated with an inverse method. For example, Ruppert et al. (2012) used a global optimization algorithm to identify the region with high posterior density and then used radial basis function to build a surrogate for the logarithm of posterior density in this local area. Zhang et al. (2013) performed a similar approach, except that the surrogate was constructed for the model responses. Li & Marzouk (2014) proposed an adaptive approach to finding a distribution that is close to the posterior distribution by minimizing the cross entropy, and then constructing a locally accurate PCE surrogate therein. Recently, Gaussian process (GP) regression (Rasmussen & Williams, 2006) was combined with MCMC simulation to iteratively refine the surrogate over the posterior distribution (Gong & Duan, 2017; Machac et al., 2018; Wang & Li, 2018; Zhang et al., 2016; Zhang et al., 2018b). It is shown that this approach can obtain satisfying inversion results with only a small number of high-fidelity model evaluations. In each iteration, to refine the surrogate locally, new high-fidelity training data will be generated based on the results of surrogate-based MCMC simulation. When selecting the new design points, different rules can be employed. One simple and effective way is to directly draw a certain number of random samples in the Markov chain(s) after convergence (Zhang et al., 2016). These samples can be viewed as random draws from the approximated posterior distribution. Machac et al. (2018) adopted a similar approach to drawing new design points for high-fidelity model evaluations, but they intentionally stretched these samples by a factor of 1.1, i.e., the new samples were actually drawn from a distribution that is 1.1 wider than the approximated posterior. Gong & Duan (2017) proposed a more elaborate method to generate representative samples from the approximated posterior, which is implemented as follows: 1) select the sample with the largest posterior density; 2) divide the converged chain states into several quintile ranges, and select in each range a new sample that has the largest distance to its nearest neighbor. Moreover, considering that the aim is to reduce the surrogate uncertainty locally, the sample that has the largest variance of GP prediction (or other criteria derived accordingly) can be selected as the new design point (Wang & Li, 2018).

Nevertheless, in most of the above approaches, approximation error of the surrogate is not explicitly quantified. In this paper, we propose two strategies to address this issue and thus improve the simulation accuracy. In the first strategy, when a surrogate has been built based on the high-fidelity training data, we further quantify its approximation uncertainty with a Bayesian method, e.g., MCMC, conditioned on the same training data set. This strategy applies to a broad class of surrogates whose predictions are largely determined by their hyperparameters (or coefficients). In implementing this strategy, we can use MCMC to characterize the uncertainty of the surrogate hyperparameters, and thus obtain the surrogate prediction uncertainty. Note, that conditioned on the training data set, GP can simultaneously provide the posterior mean and variance of the surrogate prediction (Rasmussen & Williams, 2006). Thus, it can be viewed as one special case of the first strategy. In the second strategy, we directly correct the approximation error of the primary surrogate with another surrogate (i.e., a secondary surrogate), based on the difference between the high-fidelity and the primary surrogate model outputs at the existing design points. Then we use the primary surrogate outputs plus the secondary surrogate outputs as the corrected surrogate outputs in MCMC simulations. The second strategy can be applied with almost all data-driven surrogates.

Generally, the posterior distribution of model parameters occupies only a small proportion of the prior distribution. Thus, the variance of the high-fidelity model outputs in the posterior will be smaller than that in the prior, which means that the input-output relationship of the high-fidelity model can be more accurately captured by the surrogate in the posterior than in the prior, if only a limited number of high-fidelity model evaluations are afforded. With an active learning process proposed in our previous works (Zhang et al., 2016; Zhang et al., 2018b), we can gradually reduce the surrogate approximation error to a small level in the posterior region. To our best knowledge, this adaptive approach to explicitly quantifying and gradually reducing the surrogate approximation error over the posterior distribution is rather new. And as will be demonstrated in the latter part of this paper, this adaptive approach can enjoy both high efficiency and accuracy in MCMC simulations. Thus, we believe that the proposed method has its value in both theoretical and practical aspects.

In this paper, considering that abundant published works on adaptive surrogate modeling for inverse problems have used PCE (Conrad et al., 2016; Li & Marzouk, 2014; Yan & Zhou, 2019) and/or GP (Conrad et al., 2016; Gong & Duan, 2017; Machac et al., 2018; Wang & Li, 2018; Zhang et al., 2016; Zhang et al., 2018b), so we decide to use these two popular surrogates for demonstration. The performances of other surrogates will not be evaluated here. For nonlinear problems, one has to set high PCE orders (it generally means a large number of training data) to accurately capture the input-output relationship of the high-fidelity model, which makes the full PCE rather inefficient (Xiu, 2010). Thus, we adopt a state-of-the-art PCE method, i.e., the sparse PCE based on least angle regression (LARS-PCE) proposed by Blatman & Sudret (2011), to construct the surrogate. The LARS-PCE method can adaptively detect the important PCE terms and assign zero-coefficients to the unimportant terms (which is called the sparse representation). In practical applications, one can set high PCE orders to catch the nonlinear input-output relationship of the high-fidelity model without significantly increasing the computational cost of surrogate construction.

The remainder of this paper is structured as follows. In Section 2, we present some essential knowledge about MCMC, LARS-PCE, GP, and the adaptive approach to quantifying and gradually reducing the surrogate approximation error. Then in Section 3, we demonstrate the performance of the proposed method with three case studies involving high-dimensionality, multi-modality and a real-world application, respectively. Finally, some conclusions and discussions are provided in Section 4.

## 2. Methods

For simplicity, we can represent the hydrological model of concern in the following form:

$$\tilde{\mathbf{y}} = f(\mathbf{m}) + \boldsymbol{\varepsilon}, \tag{1}$$

where $f(\cdot)$ is the high-fidelity model with input parameters $\mathbf{m} \in \mathbb{R}^{N_m \times 1}$, and $\tilde{\mathbf{y}} \in \mathbb{R}^{N_y \times 1}$ is a vector for the measurements with error $\boldsymbol{\varepsilon} \in \mathbb{R}^{N_y \times 1}$. Our initial knowledge about the model parameters $\mathbf{m}$ is represented by the prior distribution $p(\mathbf{m})$. When the measurements $\tilde{\mathbf{y}}$ are available, we can update our knowledge about $\mathbf{m}$ according to Bayes' theorem:

$$p(\mathbf{m}|\tilde{\mathbf{y}}) = \frac{p(\mathbf{m})p(\tilde{\mathbf{y}}|\mathbf{m})}{\int p(\tilde{\mathbf{y}}|\mathbf{m})p(\mathbf{m})d\mathbf{m}} \propto p(\mathbf{m})\mathcal{L}(\mathbf{m}|\tilde{\mathbf{y}}), \tag{2}$$

where $p(\mathbf{m}|\tilde{\mathbf{y}})$ is the posterior distribution, $\mathcal{L}(\mathbf{m}|\tilde{\mathbf{y}}) \equiv p(\tilde{\mathbf{y}}|\mathbf{m})$ is the likelihood function, and $p(\tilde{\mathbf{y}}) = \int p(\tilde{\mathbf{y}}|\mathbf{m})p(\mathbf{m})d\mathbf{m}$ is the so-called Bayesian evidence or marginal likelihood, which is a constant. When the measurement error $\boldsymbol{\varepsilon}$ is normally distributed with zero-mean and standard deviation $\tilde{\boldsymbol{\sigma}} = \{\tilde{\sigma}_1, \ldots, \tilde{\sigma}_{N_y}\}$, the likelihood can be expressed as:

$$\mathcal{L}(\mathbf{m}|\tilde{\mathbf{y}}) = \prod_{i=1}^{N_y} \frac{1}{\tilde{\sigma}_i \sqrt{2\pi}} \exp\left[-\frac{1}{2}\left(\frac{\tilde{y}_i - f_i(\mathbf{m})}{\tilde{\sigma}_i}\right)^2\right], \tag{3}$$

where $\tilde{y}_i$ and $f_i(\mathbf{m})$ are the *i*th components of the measurements $\tilde{\mathbf{y}}$ and the model responses $f(\mathbf{m})$, respectively. In nonlinear, complex problems, $p(\mathbf{m}|\tilde{\mathbf{y}})$ usually cannot be derived analytically. Then one has to approximate the posterior with a numerical method, e.g., MCMC.

2.1. Markov Chain Monte Carlo Simulation

MCMC works by generating a (quasi-)random walk through the parameter space to form one or more Markov chains. Under strict conditions of ergodicity and reversibility, the chain(s) can converge to the target distribution, i.e., the posterior distribution of model parameters, after enough steps (i.e., after the so-called burn-in period). Then one can use samples in the chains to approximate the statistics of $p(\mathbf{m}|\tilde{\mathbf{y}})$. MCMC can be implemented with the following procedure (Brooks et al., 2011):

Step 1. At the initial step, draw an initial state $\mathbf{m}_0$ of the Markov chain(s) from the prior distribution, i.e., $\mathbf{m}_0 \sim p(\mathbf{m})$.

Step 2. At step $t$ ($t \geq 1$), generate a candidate $\mathbf{m}_p$ from a proposal distribution $q_t(\cdot)$ conditioned on the previous state $\mathbf{m}_{t-1}$, i.e., $\mathbf{m}_p \sim q_t(\mathbf{m}_{t-1}, \cdot)$.

Step 3. According to the Metropolis-Hastings rule (Hastings, 1970; Metropolis et al., 1953), accept $\mathbf{m}_p$ with the following probability:

$$p_{\text{acc}} = \min\left[1, \frac{p(\mathbf{m}_p|\tilde{\mathbf{y}})q(\mathbf{m}_p \to \mathbf{m}_{t-1})}{p(\mathbf{m}_{t-1}|\tilde{\mathbf{y}})q(\mathbf{m}_{t-1} \to \mathbf{m}_p)}\right], \tag{4}$$

where $p(\mathbf{m}_p|\tilde{\mathbf{y}})$ and $p(\mathbf{m}_{t-1}|\tilde{\mathbf{y}})$ are the posterior densities of $\mathbf{m}_p$ and $\mathbf{m}_{t-1}$, $q(\mathbf{m}_p \to \mathbf{m}_{t-1})$ and $q(\mathbf{m}_{t-1} \to \mathbf{m}_p)$ are the probabilities of trail moves from $\mathbf{m}_p$ to $\mathbf{m}_{t-1}$ and from $\mathbf{m}_{t-1}$ to $\mathbf{m}_p$, respectively.

Step 4. Let $t = t + 1$, repeat Steps 2-3 until $t = t_{\max}$ (i.e., the user-defined maximum iteration number).

Over the past decade, many algorithms have been developed to improve the efficiency of MCMC simulations. In this paper, we adopt the DREAM(ZS) algorithm that uses a mixture of parallel and snooker proposal distributions to generate the candidate states based on a thinned history of the Markov chains. It is noted here that DREAM(ZS) uses the Metropolis rule where $q(\mathbf{m}_p \to \mathbf{m}_{t-1}) = q(\mathbf{m}_{t-1} \to \mathbf{m}_p)$, to calculate the acceptance probability, i.e.,

$$p_{\text{acc}} = \min\left[1, \frac{p(\mathbf{m}_p|\tilde{\mathbf{y}})}{p(\mathbf{m}_{t-1}|\tilde{\mathbf{y}})}\right]. \tag{5}$$

For the implementation details of DREAM(ZS), one can refer to the related works (Laloy & Vrugt, 2012; Laloy et al., 2013; Vrugt, 2016).

2.2. Surrogate-Based MCMC Simulation Considering Surrogate Approximation Error

Generally, MCMC requires a large number of iterations to reach convergence, which means a high computational cost when the hydrological model $f(\mathbf{m})$ is CPU-intensive. In this situation, using a CPU-efficient surrogate can greatly accelerate the MCMC simulation. Given a set of training data $\mathbf{Y} = \{f(\mathbf{m}_1), \ldots, f(\mathbf{m}_{N_t})\}$ evaluated at $N_t$ design points $\mathbf{M} = \{\mathbf{m}_1, \ldots, \mathbf{m}_{N_t}\}$ with the high-fidelity model, we can construct a surrogate model using a data-driven method, i.e.,

$$\hat{f}(\mathbf{m}, \boldsymbol{\phi}) \approx f(\mathbf{m}), \tag{6}$$

where $\boldsymbol{\phi}$ are the hyperparameters of the surrogate. In latter part of this paper, for simplicity, we may omit $\boldsymbol{\phi}$ and use $\hat{f}(\mathbf{m})$ to represent the surrogate.

However, the approximation error of the surrogate, i.e., $g(\mathbf{m}) = f(\mathbf{m}) - \hat{f}(\mathbf{m})$, is inevitable and can introduce some bias to the surrogate-based MCMC simulation. To address this issue, below we propose two strategies to explicitly quantify the surrogate approximation error.

1. The first strategy (Strategy A) treats the hyperparameters $\boldsymbol{\phi}$ as random variables. Conditioned on the already existing training data $\mathbf{Y}$ (which means no extra high-fidelity model evaluations are needed), we can estimate the posterior distribution of $\boldsymbol{\phi}$, i.e., $p(\boldsymbol{\phi}|\mathbf{Y})$, with a Bayesian method, e.g., MCMC. Then we can propagate the posterior uncertainty of $\boldsymbol{\phi}$ to the surrogate prediction with Monte Carlo simulation. When $Q$ posterior realizations of $\boldsymbol{\phi}$, i.e., $\{\boldsymbol{\phi}_1, \ldots, \boldsymbol{\phi}_Q\}$, are available, we can obtain $Q$ surrogate predictions, $\{\hat{f}(\mathbf{m}^*, \boldsymbol{\phi}_1), \ldots, \hat{f}(\mathbf{m}^*, \boldsymbol{\phi}_Q)\}$, at an arbitrary parameter sample $\mathbf{m}^*$. From $\{\hat{f}(\mathbf{m}^*, \boldsymbol{\phi}_1), \ldots, \hat{f}(\mathbf{m}^*, \boldsymbol{\phi}_Q)\}$, we can estimate the mean, variance and higher-order statistics (if necessary) of the surrogate prediction at $\mathbf{m}^*$. In this way, the uncertainty of the surrogate prediction can be quantified.

In the surrogate-based MCMC simulation, when the surrogate prediction uncertainty is ignored, the prediction of the surrogate is prone to be over-confident, which causes bias in the estimated posterior distribution. To incorporate the surrogate prediction uncertainty in the MCMC simulation, we can add random realizations of surrogate approximation error to surrogate mean prediction to serve as the model response used in the likelihood function. When the distribution of surrogate approximation error is close to Gaussian, we can directly add the covariance matrix of surrogate approximation error to the covariance matrix of measurement error, as the covariance of total error. Then the estimated posterior should cover a wider range that is likely to include the true posterior. Furthermore, this wider posterior indicates where to allocate new design points if we would like to refine the surrogate locally and thus to obtain a more accurate estimation.

2. The second strategy (Strategy B) uses another surrogate (in other words, a secondary surrogate) to simulate the approximation error of the primary surrogate, i.e., $\hat{g}(\mathbf{m}, \boldsymbol{\varphi}) \approx g(\mathbf{m})$, where $\boldsymbol{\varphi}$ are the hyperparameters of the secondary surrogate. Strategy B does not requires extra high-fidelity model evaluations either, as it uses $\mathbf{G} = \{f(\mathbf{m}_1) - \hat{f}(\mathbf{m}_1), \ldots, f(\mathbf{m}_{N_t}) - \hat{f}(\mathbf{m}_{N_t})\}$ as the training data for the secondary surrogate, where $\mathbf{Y} = \{f(\mathbf{m}_1), \ldots, f(\mathbf{m}_{N_t})\}$ are already available. Then we can use $\hat{f}_c(\mathbf{m}, \boldsymbol{\phi}, \boldsymbol{\varphi}) = \hat{f}(\mathbf{m}, \boldsymbol{\phi}) + \hat{g}(\mathbf{m}, \boldsymbol{\varphi})$ as the corrected surrogate in the MCMC simulation.

In this paper, two surrogate construction methods, i.e., LARS-PCE and GP are used. For Strategy A, it is easy to quantify the prediction uncertainty of PCE and GP within the Bayesian framework, and GP by its nature can also provide the posterior mean and variance of the surrogate prediction. For Strategy B, we will compare the following four combinations, i.e., PCE + GP, PCE + PCE, GP + PCE and GP + GP. Here the surrogate before the plus sign in each combination is the primary surrogate, and the surrogate after the plus sign is the secondary surrogate that simulates the approximation error of the primary surrogate.

### 2.2.1. Surrogate Construction with LARS-PCE

In this section, we will briefly introduce how to use the LARS-PCE method proposed by Blatman & Sudret (2011) to construct a surrogate. In PCE, the model output is approximated with a set of orthogonal polynomials in the following way:

$$f(\mathbf{m}) \approx \hat{f}(\mathbf{m}) = \sum_{i=0}^{P-1} c_i \psi_i(\mathbf{m}) = \mathbf{c}[\boldsymbol{\psi}(\mathbf{m})]^\mathrm{T}, \tag{7}$$

where $\boldsymbol{\psi}(\mathbf{m}) = \{\psi_0(\mathbf{m}), \ldots, \psi_{P-1}(\mathbf{m})\}$ are orthogonal polynomials over the model parameter distribution, $\mathbf{c} = \{c_0, \ldots, c_{P-1}\}$ are the corresponding coefficients, which can be determined with the intrusive Galerkin method, or the non-intrusive projection/regression method (Choi et al., 2004; Xiu, 2010), and $P$ is the total number of PCE terms. Here, "orthogonal" means that

$$\langle \psi_i(\mathbf{m}), \psi_j(\mathbf{m}) \rangle = \int_\Theta \psi_i(\mathbf{m}) \psi_j(\mathbf{m}) \, p(\mathbf{m}) d\mathbf{m} = \begin{cases} 0 & \text{if } i \neq j, \\ 1 & \text{if } i = j, \end{cases} \quad (8)$$

where $\Theta$ is the support of $p(\mathbf{m})$.

As the number of $f(\mathbf{m})$ evaluations needed by surrogate construction increases with the number of PCE terms (which grows dramatically with parameter dimension), it is beneficial to only retain the terms that have significant impacts on the model output and set zero-coefficients to the other unimportant terms. This turns the classical full PCE to a sparse PCE (Blatman & Sudret, 2008). In LARS-PCE, a hyperbolic scheme is used to truncate the PCE terms for a sparse representation, which favors the low-order interactions more than the high-order interactions in the model. The significant PCE coefficients are automatically detected by least angle regression (Efron et al., 2004) in an adaptive manner. To check the accuracy of the sparse PCE surrogate, the corrected leave-one-out error can be calculated for assessment. Then the obtained sparse PCE can be represented as:

$$f(\mathbf{m}) \approx \hat{f}(\mathbf{m}) = \sum_{j=0}^{Q-1} c_j^s \psi_j^s(\mathbf{m}) = \mathbf{c}^s [\boldsymbol{\psi}^s(\mathbf{m})]^\mathrm{T}, \quad (9)$$

where $\boldsymbol{\psi}^s(\mathbf{m}) = \{\psi_0^s(\mathbf{m}), \ldots, \psi_{Q-1}^s(\mathbf{m})\}$ are the significant orthogonal polynomials, $\mathbf{c}^s = \{c_0^s, \ldots, c_{Q-1}^s\}$ are the corresponding coefficients, and $Q$ is the total number of PCE terms that are retained.

In LARS-PCE, one can set high PCE orders for nonlinear problems to obtain good performances without significantly increasing the computational cost of surrogate construction (i.e., the number of retained PCE terms $Q$ is not large). As the LARS-PCE algorithm itself is not the focus of this paper, further details about this method are not presented here. Interested readers can refer to Blatman & Sudret (2011). Moreover, a MATLAB package named UQLab that includes LARS-PCE and many other useful algorithms for uncertainty quantification is also available online (Marelli & Sudret, 2014).

2.2.2. Surrogate Construction with GP

As a generic supervised learning method, GP uses a (multivariate) Gaussian distribution to simulate the model response:

$$\hat{f}(\cdot) \sim \mathcal{GP}(\mu(\cdot), k(\cdot, \cdot)), \quad (10)$$

where $\mu(\cdot)$ is the mean function, and $k(\cdot, \cdot)$ is the covariance function. In this paper, the zero-mean function $\mu(\mathbf{m}) = 0$ and squared exponential covariance function

$$k(\mathbf{m}, \mathbf{m}') = \sigma^2 \exp\left[-\frac{1}{2}\sum_{n=1}^{N_\mathrm{m}} \left(\frac{\mathbf{m}_n - \mathbf{m}'_n}{l_n}\right)^2\right] \tag{11}$$

are used. The zero-mean function and squared exponential covariance function are the most widely-used forms and applicable to a wide range of problems (Rasmussen & Williams, 2006). Here, $\sigma$ and $\{l_1,\ldots,l_{N_\mathrm{m}}\}$ are the hyperparameters of the covariance function, $\mathbf{m}$ and $\mathbf{m}'$ are two arbitrary sets of model parameters, respectively.

Based on the training data $\mathbf{Y} = \{f(\mathbf{m}_1), \ldots, f(\mathbf{m}_{N_\mathrm{t}})\}$ at $N_\mathrm{t}$ design points $\mathbf{M} = \{\mathbf{m}_1, \ldots, \mathbf{m}_{N_\mathrm{t}}\}$, we can train the GP surrogate by minimizing the following objective function (i.e., the negative log marginal likelihood function)

$$\mathcal{O} = \frac{1}{2}\log|\mathbf{K}| + \frac{1}{2}\mathbf{Y}^\mathrm{T}\mathbf{K}^{-1}\mathbf{Y} + \frac{N_\mathrm{t}}{2}\log 2\pi, \tag{12}$$

and obtain the conditional mean and variance of the surrogate prediction at an arbitrary set of model parameters $\mathbf{m}^*$:

$$\mu_{|\mathbf{Y}}(\mathbf{m}^*) = \mathbf{k}(\mathbf{m}^*, \mathbf{M})\mathbf{K}^{-1}\mathbf{Y}, \tag{13}$$

$$\sigma^2_{|\mathbf{Y}}(\mathbf{m}^*) = k(\mathbf{m}^*, \mathbf{m}^*) - \mathbf{k}(\mathbf{m}^*, \mathbf{M})\mathbf{K}^{-1}\mathbf{k}(\mathbf{M}, \mathbf{m}^*), \tag{14}$$

where the $i$th component of $\mathbf{k}(\mathbf{m}^*, \mathbf{M}) \in \mathbb{R}^{1 \times N_\mathrm{t}}$ is $k(\mathbf{m}^*, \mathbf{m}_i)$, $\mathbf{K} = \mathbf{k}(\mathbf{M}, \mathbf{M}) + \sigma_n^2 \mathbf{I}_{N_\mathrm{t}}$, the component at the $i$th row and $j$th column of $\mathbf{k}(\mathbf{M}, \mathbf{M}) \in \mathbb{R}^{N_\mathrm{t} \times N_\mathrm{t}}$ is $k(\mathbf{m}_i, \mathbf{m}_j)$, and $\mathbf{k}(\mathbf{M}, \mathbf{m}^*) \in \mathbb{R}^{N_\mathrm{t} \times 1}$ is the transpose of $\mathbf{k}(\mathbf{m}^*, \mathbf{M})$, $\sigma_n^2$ is the variance of possible noise in the data, and $\mathbf{I}_{N_\mathrm{t}}$ is the identity matrix of size $N_\mathrm{t}$. For more details about GP, interested readers can refer to Rasmussen & Williams (2006).

2.3. The Adaptive Approach to Reducing the Surrogate Approximation Error

Nevertheless, for many nonlinear hydrological systems, given a limited (and even a rather large) number of high-fidelity training data randomly drawn from the prior distribution, the original input-output relationship generally cannot be accurately captured even by a state-of-the-art surrogate modeling technique, e.g., LARS-PCE or GP. On the other hand, as the posterior usually occupies a small proportion of the prior distribution, the variance of the high-fidelity model outputs in the posterior will be smaller than that in the prior, which means that the input-output relationship of the high-fidelity model can be more easily captured by the surrogate in the posterior region. Based on this idea, here we formulate an active learning process that iteratively adds new design points that gradually approach to the high posterior density region to refine the surrogate locally. In each iteration, the added design points are sampled from the approximated posterior from the surrogate-based MCMC simulation. After enough iterations, we can obtain a locally accurate surrogate and an accurate estimation of the posterior distribution.

A complete procedure of the approach proposed in this paper is given in Algorithm 1. At the beginning, $N_\mathrm{ini}$ random samples are drawn from the prior distribution of model parameters, i.e., $\mathbf{M} = \{\mathbf{m}_1,\ldots,\mathbf{m}_{N_{\text{ini}}}\}$. With $\mathbf{M}$ and the corresponding high-fidelity model outputs $\mathbf{Y} = \{f(\mathbf{m}_1),\ldots,f(\mathbf{m}_{N_{\text{ini}}})\}$, we can train and obtain an initial surrogate $\hat{f}_0(\mathbf{m})$ using an adequate data-driven method. Generally, the accuracy of $\hat{f}_0(\mathbf{m})$ is far from enough to obtain satisfying inversion results in MCMC simulation, even when Strategy A or Strategy B is used to account for the surrogate approximation error. Nevertheless, the approximated posterior $\tilde{p}_0(\mathbf{m}|\tilde{\mathbf{y}})$ should be slightly closer to the real posterior than the prior $p(\mathbf{m})$. From $\tilde{p}_0(\mathbf{m}|\tilde{\mathbf{y}})$, we can draw $N_a$ new design points, $\mathbf{M}_a = \{\mathbf{m}_{a,1},\ldots,\mathbf{m}_{a,N_a}\}$, that are expected to be closer to the posterior region than the design points randomly drawn from the prior distribution. Then we add them to the pool of existing training data set, i.e., $\mathbf{M} = \{\mathbf{M}, \mathbf{M}_a\}$ and $\mathbf{Y} = \{\mathbf{Y}, \mathbf{Y}_a\}$, where $\mathbf{Y}_a$ are the high-fidelity model outputs of $\mathbf{M}_a$, and construct a new surrogate $\hat{f}_1(\mathbf{m})$ accrodingly. In constructing the surrogate, the computational time increases with the number of design points. So in practice we can also use a subset of $\mathbf{Y}$, where some samples that are far from the measurements $\tilde{\mathbf{y}}$ are discarded (e.g., using the likelihood function defined in equation (3) as a measure). To improve the efficiency of MCMC simulations, we will use a thinned chain history of the previous MCMC simulation to generate initial states for the present MCMC simulation. In this way, the present MCMC simulation is actually connected to the previous one and thus it can converge to the stationary regime quickly. The aforementioned surrogate refinement and surrogate-based MCMC simulation that considers surrogate approximation error will be further iterated $I_{\max} - 1$ times. Finally, the posterior is approximated as $\tilde{p}_{I_{\max}}(\mathbf{m}|\tilde{\mathbf{y}})$.

---

**Algorithm 1** Adaptive surrogate-based MCMC simulation considering surrogate approximation error

---

1. Draw $N_{\text{ini}}$ random samples from $p(\mathbf{m})$, i.e., $\mathbf{M} = \{\mathbf{m}_1,\ldots,\mathbf{m}_{N_{\text{ini}}}\}$, calculate $\mathbf{Y} = \{f(\mathbf{m}_1),\ldots,f(\mathbf{m}_{N_{\text{ini}}})\}$.
2. Build an initial primary surrogate $\hat{f}_0(\mathbf{m})$ conditioned on $\mathbf{M}$ and $\mathbf{Y}$, or their subsets.
3. Run MCMC with $\hat{f}_0(\mathbf{m})$ using Strategy A or B, obtain $\tilde{p}_0(\mathbf{m}|\tilde{\mathbf{y}})$.
4. **for** $i = 1,\ldots,I_{\max}$ **do**

    Draw $N_a$ random samples, $\mathbf{M}_a = \{\mathbf{m}_{a,1},\ldots,\mathbf{m}_{a,N_a}\}$, from $\tilde{p}_{i-1}(\mathbf{m}|\tilde{\mathbf{y}})$, calculate $\mathbf{Y}_a = \{f(\mathbf{m}_{a,1}),\ldots,f(\mathbf{m}_{a,N_a})\}$. Let $\mathbf{M} = \{\mathbf{M}, \mathbf{M}_a\}$ and $\mathbf{Y} = \{\mathbf{Y}, \mathbf{Y}_a\}$.

    Build a new primary surrogate $\hat{f}_i(\mathbf{m})$ conditioned on $\mathbf{M}$ and $\mathbf{Y}$, or their subsets.

    Based on previous MCMC simulation results, run MCMC with $\hat{f}_i(\mathbf{m})$ using Strategy A or B, obtain $\tilde{p}_i(\mathbf{m}|\tilde{\mathbf{y}})$.

    **end for**
5. The posterior is approximated as $\tilde{p}_{I_{\max}}(\mathbf{m}|\tilde{\mathbf{y}})$.

---

It is noted that parameter sensitivity plays an important role in many situations. In the worst (unlikely) case, if all parameters are insensitive to the measurements, any inverse modeling method will perform badly. In many cases, however, there will always be some parameters that are sensitive and we can reduce their uncertainties by conditioning on the measurements (or it will be useless to collect the measurement data to calibrate the model). Then the posterior region will be narrower than the prior region, i.e., the posterior is a subset of the prior. It is understandable that constructing a locally accurate surrogate (i.e., in the subset region) is less computationally demanding than constructing a globally accurate one. To address the possible problem of parameter insensitivity, we can conduct sensitivity analysis before parameter estimation, and design the optimal collection of measurement data that contain enough information about the parameters, see, e.g., (Siade et al., 2017; Wang et al., 2018; Zhang et al., 2015).

**3. Illustrative Examples**

3.1. Example 1: A Contaminant Source Identification Problem

3.1.1. The Case with Perfect Prior Knowledge about the Conductivity Field

To demonstrate the performance of the proposed method, we will test in this section a numerical case that solves a groundwater contaminant source identification problem, which is adopted from our previous study (Zhang et al., 2018b). This case study involves nonlinear processes that have 28 unknown model parameters, which poses a challenge for solving an inverse problem.

[Figure 1 NEAR HERE]

Here, we consider steady-state saturated groundwater flow and solute transport in a confined aquifer. In the horizontal direction, the length of the domain is 20[L], and in the vertical direction, the length is 10[L] (Figure 1). On the upper and lower sides of the domain, the no-flow condition is prescribed. On the left and right sides, the constant-head condition is prescribed (12[L] and 11[L], respectively). At the initial time, the hydraulic head is 11[L] everywhere except for the left boundary (12[L]). With the above initial and boundary conditions, we can obtain a steady-state flow field by solving

$$\frac{\partial}{\partial x_i}\left(K_i \frac{\partial h}{\partial x_i}\right) = 0 \qquad (15)$$

numerically with MODFLOW (Harbaugh et al., 2000), and the velocity of pore water in the $i$th direction $v_i[\text{LT}^{-1}]$ can be obtained as follows:

$$v_i = -\frac{K_i}{\theta}\frac{\partial h}{\partial x_i}. \qquad (16)$$

Here, $x_i[\text{L}]$ signifies the distance in the respective direction ($i = 1$ for the $x$ direction and $i = 2$ for the $y$ direction), $h[\text{L}]$ is the hydraulic head, $K_i[\text{LT}^{-1}]$ represents the component of hydraulic conductivity in the $i$th direction, and $\theta[\text{-}]$ is the porosity of the aquifer, respectively. Here, the conductivity field is unknown and heterogeneous, whose log-transformed values $Y = \ln K$ are spatially correlated according to the following function:

$$C_Y(x_a, y_a; x_b, y_b) = \sigma_Y^2 \exp\left(-\frac{|x_a - x_b|}{l_x} - \frac{|y_a - y_b|}{l_y}\right), \qquad (17)$$

where $(x_a, y_a)$ and $(x_b, y_b)$ are two arbitrary locations in the domain, $\sigma_Y^2$ signifies the variance of the $Y$ field, and $l_x(l_y)$ denotes the correlation length in the $x(y)$ direction, respectively. Utilizing this correlation relationship, we can reduce the dimensionality of the $Y$ field with the Karhunen-Loève (KL) expansion (Zhang & Lu, 2004) from the model grid number (81 × 41) to the number of truncated KL terms ($N_{\mathrm{KL}}$):

$$Y(\mathbf{x}) \approx \bar{Y}(\mathbf{x}) + \sum_{i=1}^{N_{\mathrm{KL}}} \sqrt{\tau_i} s_i(\mathbf{x}) \xi_i, \tag{18}$$

where $\mathbf{x} = (x, y)$ signifies the location, $\bar{Y}(\mathbf{x})$ is the mean function of the $Y$ field, $s_i(\mathbf{x})$ and $\tau_i$ denote the eigenfunctions and eigenvalues of the correlation function of equation (17), and $\xi_i$ represent independent standard Gaussian random variables, for $i = 1, \ldots, N_{\mathrm{KL}}$. Here, some parameters are assumed to be known from field survey, i.e., $\theta = 0.25[-]$, $\sigma_Y^2 = 0.4$, $l_x = 10[\mathrm{L}]$, $l_y = 5[\mathrm{L}]$, and $\bar{Y}(\mathbf{x}) = 2$, respectively. In the flow process, the uncertainty only stems from the conductivity field, which is characterized by $N_{\mathrm{KL}} = 20$ unknown KL terms that keeps about 88.3% of the total variance of the $Y$ field.

In the flow field, some amount of non-reactive contaminant is released from an unknown point source (S) that is located somewhere in the red dashed rectangle depicted in Figure 1. The strength of the source measured by mass loading rate varies with time and is characterized by six parameters, i.e., $s_i[\mathrm{MT}^{-1}]$ during $t = i{:}i + 1[\mathrm{T}]$, for $i = 1, \ldots, 6$. Then the contaminant concentration $C[\mathrm{ML}^{-3}]$ at any time and location can be obtained by solving the advection-dispersion equation

$$\frac{\partial(\theta C)}{\partial t} = \frac{\partial}{\partial x_i}\left(\theta D_{ij}\frac{\partial C}{\partial x_j}\right) - \frac{\partial}{\partial x_i}(\theta v_i C) + q_a C_s \tag{19}$$

with MT3DMS (Zheng & Wang, 1999). Here, $t[\mathrm{T}]$ is the time, $q_a[\mathrm{T}^{-1}]$ is the volumetric flow rate per unit volume of the aquifer, $C_s[\mathrm{ML}^{-3}]$ is the concentration of the contaminant source, and $D_{ij}[\mathrm{L}^2\mathrm{T}^{-1}]$ are the hydrodynamic dispersion coefficient tensors that have the following four components:

$$\begin{cases} D_{11} = \dfrac{1}{\|\mathbf{v}\|^2}(\alpha_{\mathrm{L}} v_1^2 + \alpha_{\mathrm{T}} v_2^2), \\ D_{22} = \dfrac{1}{\|\mathbf{v}\|^2}(\alpha_{\mathrm{L}} v_2^2 + \alpha_{\mathrm{T}} v_1^2), \\ D_{12} = D_{21} = \dfrac{1}{\|\mathbf{v}\|^2}(\alpha_{\mathrm{L}} - \alpha_{\mathrm{T}}) v_1 v_2, \end{cases} \tag{20}$$

where $\alpha_{\mathrm{L}}$ and $\alpha_{\mathrm{T}}[\mathrm{L}]$ signify the longitudinal and transverse dispersivities, and $\|\mathbf{v}\|^2 = \sqrt{v_1^2 + v_2^2}$ denotes the magnitude of the velocity $\mathbf{v}$, respectively. In the solute transport process, the unknown parameters are the location of the contaminant source, i.e., $\{x_s, y_s\}[\mathrm{L}]$, and the six source strength parameters, i.e., $\{s_1, \ldots, s_6\}[\mathrm{MT}^{-1}]$. While other parameters are assumed to be known as $\alpha_{\mathrm{L}} = 0.3[\mathrm{L}]$ and $\alpha_{\mathrm{T}} = 0.03[\mathrm{L}]$, respectively.

So, in total there are 28 unknown parameters, i.e., the 20 KL terms for the conductivity field and the eight source parameters. To infer the unknown parameters, we collect measurements of hydraulic head and concentration at 15 wells denoted by the blue circles in Figure 1. As the flow field is steady-state, the head measurements are collected only once. Thus, there are 15 head measurements in total. The concentration measurements are obtained at five moments of $t = 6, 8, 10, 12, 14$[T]. Thus, there are 75 concentration measurements altogether. Here, the measurements are generated from one set of true model parameters $\mathbf{m}_{\text{true}}$ with additive errors. In this case study, the measurement error of head is assumed to fit $\varepsilon_h \sim \mathcal{N}(0, 0.01^2)$, and for concentration measurement the error fits $\varepsilon_C \sim \mathcal{N}(0, 0.01^2)$. As mentioned above, the prior distribution for each component of the 20 KL terms is a standard Gaussian distribution, i.e., $\xi_i \sim \mathcal{N}(0, 1^2)$, for $i = 1,\ldots,20$. The prior distributions for the eight source parameters are all uniform. The corresponding ranges and true values of the source parameters are listed in Table 1, and the true $Y$ field that is generated from 20 KL terms is depicted in Figure 6a.

[Table 1 NEAR HERE]

Conditioned on the 90 measurements, we implement the following different MCMC simulations to infer the 28 unknown parameters. These approaches include (1) the high-fidelity MCMC simulation (i.e., $f(\mathbf{m})$-MCMC) that produces the reference results, (2) the adaptive $\hat{f}(\mathbf{m})$-MCMC simulation using PCE without error estimation, (3) the adaptive $\hat{f}(\mathbf{m})$-MCMC simulation using GP without error estimation, (4) the adaptive $\hat{f}(\mathbf{m})$-MCMC simulation that uses Strategy A to account for surrogate approximation error of PCE, (5) the adaptive $\hat{f}(\mathbf{m})$-MCMC simulation that uses Strategy A to account for surrogate approximation error of GP, (6) the adaptive $\hat{f}(\mathbf{m})$-MCMC simulation adopting Strategy B (PCE + GP), (7) the adaptive $\hat{f}(\mathbf{m})$-MCMC simulation adopting Strategy B (PCE + PCE), (8) the adaptive $\hat{f}(\mathbf{m})$-MCMC simulation adopting Strategy B (GP + PCE), and (9) the adaptive $\hat{f}(\mathbf{m})$-MCMC simulation adopting Strategy B (GP + GP), respectively. Here the adaptive $\hat{f}(\mathbf{m})$-MCMC simulation using PCE or GP without error estimation is implemented in the same way as described in Algorithm 1, except that neither Strategy A nor Strategy B is utilized to quantify the surrogate approximation error. Some brief descriptions of different adaptive surrogate-based MCMC approaches are provided in Table 2.

[Table 2 NEAR HERE]

[Figure 2 NEAR HERE]

In the high-fidelity MCMC simulation (i.e., $f(\mathbf{m})$-MCMC), there are 15 parallel chains. The length of each chain is 60,000, which means that the total number of high-fidelity model evaluations is 900,000. The Gaussian likelihood function defined in equation (3) is used to evaluate the goodness-of-fit between the measurements $\tilde{\mathbf{y}}$ and the corresponding model outputs $f(\mathbf{m})$. As shown in Figure 2, the chains (blue dots) for the eight source parameters can accurately catch the true values (black crosses). To monitor the convergence of the Markov chains, we depict the evolution of the $\hat{R}$-statistics proposed by Gelman & Rubin (1992) for the 28 model parameters in Figure

S1 (in the Supporting Information). When the values of $\hat{R}$-statistics are below 1.2 (black dashed line), we can declare a good convergence for the high-fidelity MCMC simulation. Here we use the results of the high-fidelity MCMC simulation as the reference to check the performances of different adaptive surrogate-based MCMC simulations.

[Figure 3 NEAR HERE]

In the adaptive surrogate-based MCMC simulations, the initial number of design points drawn from the prior distribution for high-fidelity model evaluation is $N_{\text{ini}} = 200$. Based on these initial training data, we can construct a surrogate to accelerate the MCMC simulation. In each iteration, we draw $N_a = 20$ random samples from the approximated posterior distribution as the new design points $\mathbf{M}_a$. Then we calculate the model outputs with the high-fidelity model and use the $N_a$ new training data to refine the surrogate locally. The iterations of surrogate refinement and surrogate-based MCMC simulation are implemented $I_{\text{max}} = 25$ times (please refer to Algorithm 1). So, in the adaptive surrogate-based MCMC simulations, the total numbers of high-fidelity model evaluations are $N_{\text{ini}} + N_a*I_{\text{max}} = 700$.

[Figure 4 NEAR HERE]

In Figure 3, we present the initial and added design points (blue dots) generated in the adaptive $\hat{f}(\mathbf{m})$-MCMC simulation using PCE without error estimation for the eight contaminant source parameters. It is clear that this approach cannot accurately capture the true parameter values (black crosses). When considering the surrogate prediction uncertainty of PCE with Strategy A, and then implement the adaptive surrogate-based MCMC simulation, an improved performance can be observed from Figure 4. Nevertheless, this performance is still not satisfying, as the uncertainty of the design points in the last few iterations is still large, compared to the chain trajectories of the same parameters obtained by the high-fidelity MCMC simulation (blue dots, Figure 2). Then we try to use GP to correct the surrogate approximation error of PCE and apply the corrected surrogate in the MCMC simulation. After the same number of iterations, this new approach can obtain a much better performance. As shown in Figure 5, the gradually added design points in the adaptive $\hat{f}(\mathbf{m})$-MCMC simulation using Strategy B (PCE + GP) can accurately identify the contaminant source parameters. In Figure 2, we also show the posterior mean estimates of the eight source parameters obtained from the above three adaptive $\hat{f}(\mathbf{m})$-MCMC simulations. From this figure, we can further conclude that neglecting the surrogate approximation error will deteriorate the active learning process that is essential to obtain a locally accurate surrogate.

[Figure 5 NEAR HERE]

We think the reasons for the improved performance introduced by Strategy B (PCE + GP) are as follows. When applying Strategy A to PCE, we are trying to add some uncertainty to the surrogate predictions so that they are more likely to cover the corresponding high-fidelity outputs. Although Strategy A can reduce the bias introduced by the surrogate approximation error (this can be concluded by comparing Figures 3 and 4), the mean estimates of PCE are still not corrected, so that the improvement is limited. When applying Strategy B (PCE + GP), we are trying to correct the mean estimates of the PCE predictions with GP directly, which can reduce more errors than Strategy A. Moreover, as indicated by Schobi et al. (2015), PCE is better at simulating the global behavior of the high-fidelity model, while GP is more skilled in capturing the local variations (e.g., the surrogate approximation error in the high posterior density region). So, we think one key factor of Strategy B (PCE + GP) is that it integrates two methods (i.e., PCE and GP) that complement each other.

[Figure 6 NEAR HERE]

Next, Figures 6b-6c depict the posterior mean estimates of $Y$ field obtained from the $f(\mathbf{m})$-MCMC simulation and the adaptive $\hat{f}(\mathbf{m})$-MCMC simulation using Strategy B (PCE + GP), respectively. The two mean fields are both derived by averaging 10,000 posterior realizations of the log-conductivity field. Compared to the true field as shown in Figure 6a, the two approaches can illuminate correctly the areas of high and low conductivity in the flow domain, yet the spatial extent is slightly underestimated. These findings indicate that the measurement data of hydraulic head and concentration contain insufficient information about the conductivity field.

[Figure 7 NEAR HERE]

In Figure 7, we further show the evolution of the error of the surrogate in the $\hat{f}(\mathbf{m})$-MCMC simulation using Strategy B (PCE + GP) at the true model parameters $\mathbf{m}_{\text{true}}$ with the following indicator:

$$\text{Err}(i) = \sqrt{\frac{1}{N_y} \sum_{j=1}^{N_y} \left[\frac{f_j(\mathbf{m}_{\text{true}}) - \hat{f}_{i,j}(\mathbf{m}_{\text{true}})}{\tilde{\sigma}_j}\right]^2}, \quad (21)$$

where $i$ is the iteration index, $\tilde{\sigma}_j$ is the standard deviation of the $j$th measurement error, $\hat{f}_{i,j}(\mathbf{m}_{\text{true}})$ and $f_j(\mathbf{m}_{\text{true}})$ are the $j$th surrogate output (PCE + GP) constructed in the $i$th iteration and the $j$th high-fidelity output, respectively. It is clear that by combining active learning and Strategy B (PCE + GP) in the surrogate-based MCMC simulation, we can gradually build a very accurate surrogate near the true model parameters.

In our previous paper (Zhang et al., 2016), we have proposed an adaptive GP-based MCMC algorithm to efficiently estimate groundwater model parameters. This algorithm is essentially the same as the adaptive $\hat{f}(\mathbf{m})$-MCMC approach that uses Strategy A to account for surrogate approximation error of GP. In Figure 8, we plot the initial and added design points generated in this approach. Compared to the similar approach that uses PCE (see Figure 4), the approach that uses GP exhibits a much better performance. The reason might be as follows. PCE tries to approximate the global behavior of the high-fidelity model using a set of orthogonal polynomials. Thus, adding some new training data in the high posterior density region cannot significantly improve the surrogate performance in this local area. In other words, the local accuracy of the surrogate that is expected to be improved by the new training data will be neutralized by training data that are far from this area. For GP, however, adding some new training data in the high posterior density region can significantly improve the surrogate accuracy locally, as this surrogate works by interpolating the high-fidelity outputs as a function of neighboring design points. Nevertheless, as discussed earlier, using GP to remedy the weakness of PCE can also bring a satisfying performance, as it combines the global strength of PCE and local ability of GP.

[Figure 8 NEAR HERE]

Then one question arises, i.e., in the adaptive surrogate-based MCMC simulation, whether using two surrogates (PCE and GP) that can complement each other, will outperform only using GP that considers the prediction uncertainty? In Figure 9, we compare the marginal posterior distributions of the eight contaminant source parameters estimated by the $f(\mathbf{m})$-MCMC simulation (reference results, green curves), the adaptive $\hat{f}(\mathbf{m})$-MCMC simulation using Strategy A (GP, red dashed curves), and the adaptive $\hat{f}(\mathbf{m})$-MCMC simulation using Strategy B (PCE + GP, blue dashed curves). These marginal density curves are estimated from 90,000 posterior samples obtained by each MCMC approach using a normal kernel function. If and only if the results of a proposed method agree with the reference results, we can regard this method as working properly. As shown in Figure 9, the results of Strategy A (GP, red dashed curves) exhibit considerable deviations from the reference results (green curves), especially for $x_s$, $y_s$, and $s_3$-$s_6$. If we adopt a more elaborate method to generate new design points for the local refinement of GP surrogate, e.g., the ASMO-PODE method (Gong & Duan, 2017), we should be able to improve the performance of Strategy A (GP), e.g., see the magenta dash-dot curves in Figure 9 and the trace plots of design points for the eight source parameters in Figure S2. The above results indicate that strategies that have different purposes, e.g., quantifying surrogate error or improving representativeness of new design points, can both improve the performance of adaptive surrogate-based MCMC simulations.

[Figure 9 NEAR HERE]

Nevertheless, one merit of the adaptive $\hat{f}(\mathbf{m})$-MCMC approach using Strategy A (GP) is that, it needs fewer iterations to reach convergence. For example, in Figure S3, when $N_{\text{ini}} = 50$, $N_a = 6$, and $I_{\max} = 20$, the adaptive $\hat{f}(\mathbf{m})$-MCMC simulation using Strategy A (GP) can still maintain a rather good performance. However, if the approximation error of GP is ignored, with the same number (170) of high-fidelity model evaluations, the performance of the adaptive GP-based MCMC simulation will significantly deteriorate, as shown in Figure S4. Besides, if we build a secondary GP surrogate for the approximation error of the primary GP surrogate, we can observe some improvement (see Figure S5) compared to the approach ignoring GP surrogate error (Figure S4), but the performance as shown in Figure S5 is still not as good as that in Figure S3. This is caused by the fact that the GP surrogate is actually a method of interpolation, such that the surrogate approximation errors at the existing design points $\{\mathbf{m}_1, \ldots, \mathbf{m}_{N_t}\}$ are rather small. When the training data are noise free, the GP approximation errors at existing design points should be zero. In practice, we use a general GP code that considers possible noise in the training data. When we use noise free training data, the approximation errors of the GP surrogate at the design points will be very small but not necessarily zero. This treatment of data noise is default in most popular GP codes, as it is more realistic in that the observed values at the design points are not the exact system outputs, but a noisy realization of them (even the high-fidelity model contains model errors). Based on the near-zero training data of $\{f(\mathbf{m}_1) - \hat{f}(\mathbf{m}_1), \ldots, f(\mathbf{m}_{N_t}) - \hat{f}(\mathbf{m}_{N_t})\}$, the secondary surrogate may significantly underestimate the error of the primary GP prediction at an arbitrary new parameter sample other than the design points.

In the Supporting Information, we also compare two other adaptive surrogate-based MCMC simulations that consider surrogate approximation errors with Strategy B. By comparing Figure S6 with Figure 3, we can see that simulating the approximation error of PCE with a secondary PCE surrogate can bring some improvement in the performance. From Figure S7, however, using PCE to account for the approximation error of GP will make the final results slightly biased, especially for $x_s$ and $y_s$. Thus, based on the above comparisons, we recommend to use the adaptive $\hat{f}(\mathbf{m})$-MCMC approach that uses Strategy B (PCE + GP) (note, not GP + PCE) to account for the surrogate approximation error.

In the proposed method, the surrogate-based MCMC simulation is not implemented only once, but multiple times. During each MCMC simulation, the surrogate model used is unchanged. When the MCMC simulation at the $i$th step is finished, we can obtain an approximated posterior. Then we draw some parameter samples from the approximated posterior and calculate their high-fidelity model outputs as new train data to refine the surrogate locally. At step $i+1$, using this updated surrogate, we restart a new MCMC simulation that has its own burn-in period. To improve the search efficiency, starting states of this new MCMC simulation are generated from the previous MCMC simulation results. Moreover, the proposed $\hat{f}(\mathbf{m})$-MCMC approach using Strategy B (PCE + GP) can obtain almost identical marginal posterior distributions of model parameters as the high-fidelity MCMC simulation (see Figure 9). Thus, the proposed approach that adaptively refines the surrogate over the posterior distribution does not violate the theory of MCMC and can work properly.

3.1.2. The Case with Imperfect Prior Knowledge about the Conductivity Field

In the above simulations, the true field (Figure 6a) is generated from 20 KL terms. By doing so, we can test the proposed method in an ideal condition, i.e., the error only originates from the measurement process. Nevertheless, the KL expansion, which is a dimension reduction method, introduces structural error to the high-fidelity model. Below we will consider a more realistic scenario where the true field is generated from sequential Gaussian simulation using the Geostatistical Simulation Library (GSLIB) (Deutsch & Journel, 1995). In this way, the true field will have a much higher dimension, which is equal to the number of model grids (81 × 41). Here, the mean, variance, correlation lengths in the $x$ and $y$ direction of the $Y$ field are $\bar{Y}(\mathbf{x}) = 2$, $\sigma_Y^2 = 0.4$, $l_x = 10[\mathrm{L}]$, and $l_y = 5[\mathrm{L}]$, respectively. To introduce some extra model structural error, a different correlation function,

$$C_Y(x_a, y_a; x_b, y_b) = \sigma_Y^2 \exp\left(-\sqrt{\frac{(x_a-x_b)^2}{l_x^2} + \frac{(y_a-y_b)^2}{l_y^2}}\right), \qquad (22)$$

is used in sequential Gaussian simulation to generate the true field. As shown in Figure 11a, the new $Y$ field exhibits more complex patterns than the one in Figure 6a. The true values of the contaminant source parameters are listed in Table 1. Then we generate the measurements by running the high-fidelity model with the true parameters and adding random errors that fit $\varepsilon_h \sim \mathcal{N}(0, 0.01^2)$ and $\varepsilon_C \sim \mathcal{N}(0, 0.01^2)$.

[Figure 10 NEAR HERE]

When inferring the unknown parameters, we still use equation (17) as the correlation function of the $Y$ field, and 20 truncated KL terms for a reduced-dimension representation in the high-fidelity model. This represents the epistemic error in our prior knowledge about the hydrological system. Thus, there are still 28 parameters to be estimated, whose prior distributions are the same as in the simulations conducted in Section 3.1.1. In the high-fidelity MCMC simulation, there are 15 parallel chains. The length of each chain is 60,000, which means that the total number of high-fidelity model evaluations is 900,000. In the adaptive $\hat{f}(\mathbf{m})$-MCMC approach using Strategy B (PCE + GP), the initial number of high-fidelity training data is $N_{\text{ini}}=200$. In each iteration, $N_a=20$ new design points are added to refine the surrogate locally. The process of surrogate-based MCMC simulation and surrogate refinement is iterated $I_{\text{max}}=40$ times. Thus, the total number of high-fidelity model evaluations is 1,000. As shown in Figure 10, the high-fidelity MCMC simulation (blue dots) takes a much longer time than that in Figure 2 to move close to the true values (black crosses) of the contaminant source parameters. Moreover, the estimation results of the high-fidelity MCMC simulation and the adaptive surrogate-based MCMC simulation (posterior mean estimates represented by red squares) slightly deviate from the true values, especially for $y_s$ and $s_2$. In Figures 11b and 11c, we depict the mean estimates of $Y$ field from the high-fidelity MCMC simulation and the adaptive surrogate-based MCMC simulation, respectively. Although using dimension reduction with the KL expansion in parameter estimation will lose some details about the $Y$ field, the two approaches can still capture the main patterns of the true field, and the surrogate-based approach has a slightly better performance.

[Figure 11 NEAR HERE]

3.2. Example 2: Contaminant Source Identification with Multimodal Posterior

To further demonstrate the performance of the proposed method, below we will test a new inverse problem with multimodal posterior distribution. Here, we consider a homogeneous conductivity field with a known value of $K = 8[\text{LT}^{-1}]$. The contaminant source is characterized by five parameters, i.e., the location $(x_s, y_s)$, the strength $(S_s)$, the start time of release $(t_{\text{on}})$ and the end time of release $(t_{\text{off}})$. In this case, the unknown parameters are the five contaminant source parameters, whose prior ranges and true values are given in Table 3. To infer the unknown parameters, we collect concentration measurements at $t = 6, 8, 10, 12, 14$[T] from a well denoted by the blue square in Figure 1. The measurement error fits $\varepsilon_C \sim \mathcal{N}(0, 0.01^2)$. Other settings are the same as the cases tested in Section 3.1. Then we compare the posterior distributions estimated by the high-fidelity MCMC simulation (the chain number is 10 and the length of each chain is 4,000) and the adaptive surrogate-based MCMC simulation using Strategy B (PCE + GP, $N_{\text{ini}}$=40, $N_a$=10, and $I_{\max}$=10). As shown in Figure 12, the two approaches can obtain almost identical posterior estimations of the five unknown parameters, and the multimodal posterior distribution of $y_s$ can be well identified.

[Figure 12 NEAR HERE]

[Table 3 NEAR HERE]

3.3. Example 3: Parameter Estimation for A Real-World, Large-Scale Experiment

In the above two sections, we set up a few synthetic experiments to test the proposed method. Below we will apply the proposed method to a large-scale, real-world experiment to study the mechanical, hydrological and biochemical behaviors of a high food waste content municipal solid waste landfill (Zhan et al., 2017a; Zhan et al., 2017b).

The experiment was conducted at Zhejiang University, Hangzhou, China, from June 14, 2013, and had been operated for about two years. During this period, three circles of leachate drawdown and refill were implemented. In this case, we only consider the first drawdown period (about 200 hours). The size of the bioreactor built in this experiment is 5.0m × 5.0m × 7.5m (length×width×depth), where the thickness of the waste is 5.3m. The waste pile in the bioreactor has five distinct layers with equal thickness. In each layer, the porous medium can be viewed as homogeneous.

In this bioreactor, the coupled liquid-gas flow can be described by the following equations (Helmig, 1997):

$$\phi S_w \rho_w \frac{d\rho_w}{dp_w} \frac{\partial p_w}{\partial t} - \phi \rho_w \frac{\partial S_g}{\partial t} + S_w \rho_w \frac{\partial \phi}{\partial t} \\ - \nabla \left[ \rho_w \frac{k_{rw} k_{\text{ini}}}{\mu_w} (\nabla p_w + \rho_w g) \right] - \rho_w q_w = 0, \quad (23)$$

$$\phi S_g \frac{d\rho_g}{dp_g} \frac{\partial p_w}{\partial t} - \phi S_g \frac{d\rho_g}{dp_g} \frac{dp_c}{dS_w} \frac{\partial S_g}{\partial t} + \phi \rho_g \frac{\partial S_g}{\partial t} + S_g \rho_g \frac{\partial \phi}{\partial t} \\ - \nabla \left[ \rho_g \frac{k_{rg} k_{\text{ini}}}{\mu_g} (\nabla p_w + \nabla p_c + \rho_g g) \right] - \rho_g q_g = 0, \quad (24)$$

where $\phi$ is the porosity of the porous medium [-]; $S_w$ and $S_g = 1 - S_w$ are the saturations [-], $\rho_w$ and $\rho_g$ are the densities [ML$^{-3}$], $p_w$ and $p_g$ are the pressures [ML$^{-1}$T$^{-2}$], $q_w$ and $q_g$ are the source/sink volume fluxes [L$^3$L$^{-3}$T$^{-1}$], $k_{rw}$ and $k_{rg}$ are the relative permeabilities [-], $\mu_w$ and $\mu_g$ are the viscosities [ML$^{-1}$T$^{-1}$] of the liquid (w) and gas (g) phases, respectively; $p_c = p_g - p_w$ is the capillary pressure [ML$^{-1}$T$^{-2}$]; $t$ is the time [T]; $k_{\text{ini}}$ is the intrinsic permeability of the media [L$^2$]; $g$ is the gravitational constant [LT$^{-2}$]; $\nabla$ is the del operator. Here, we use the Van

Genuchten-Mualem (VGM) model to describe the capillary pressure $p_c$, and the relative permeabilities $K_{rw}$ and $K_{rg}$ (Helmig, 1997; Mualem, 1976; Van Genuchten, 1980):

$$p_c = p_e \left(S_e^{1/m} - 1\right)^{1-m}, \tag{25}$$

$$K_{rw} = S_e^l \left[1 - \left(1 - S_e^{1/m}\right)^m\right]^2, \tag{26}$$

$$K_{rg} = (1 - S_e)^\gamma \left[1 - S_e^{1/m}\right]^{2m}, \tag{27}$$

where $p_e$ is the gas entry pressure $[ML^{-1}T^{-2}]$; $l$ and $\gamma$ are the pore-connectivity parameters [-] with suggested values of 0.50 and 0.33, respectively; $m$ is an empirical shape parameter [-]; $S_e$ is the effective saturation:

$$S_e = \frac{S_w - S_{wr}}{1 - S_{wr} - S_{gr}}, \tag{28}$$

where $S_{wr}$ and $S_{gr}$ are the residual saturations [-] of the water and gas phases, respectively.

During the first drawdown period, the bottom of the bioreactor is open to the atmosphere, so we set the gas pressure at the lower boundary as 101,325Pa, while the gas pressure at the upper boundary is 99,954Pa. The capillary pressures at the lower and upper boundaries are 0Pa and -36,346Pa, respectively. At the initial time, the gas pressure and capillary pressure at other places are assumed to vary linearly across the domain, and some parameters are obtained from experiments or literature as $k_{ini} = 1 \times 10^{-12} m^2$, $g = 9.81 m/s^2$, $p_e = 880 Pa$, $m = 0.37$, $\rho_w = 1 \times 10^3 kg/m^3$, $\rho_g = 1.24 kg/m^3$, $\mu_w = 1 \times 10^{-3} Pa \cdot s$, and $\mu_g = 1.2 \times 10^{-5} Pa \cdot s$, respectively. Here, the unknown parameters are the initial porosities of the five layers, whose prior distributions fit $\phi_i \sim \mathcal{U}(0.2, 0.8)$, for $i = 1,\ldots,5$. It should be noted here that the porosity changes over time and the intrinsic permeability will change accordingly, which can be described by the Kozeny-Carman equation (Chapuis & Aubertin, 2003). In this paper, we use an open-source simulator OpenGeoSys (Kolditz et al., 2012) to solve the above governing equations, which takes about 100 seconds for a single run. For more details about the experiment and the mathematical modeling, one can refer to (Helmig, 1997; Pinder & Gray, 2008; Zhan et al., 2017a; Zhan et al., 2017b).

To infer the five unknown model parameters, leachate flux data during the first drawdown period were collected at bottom of the reactor, which are plotted in both Figures 14a and 14b with blue circles. Then we implement the high-fidelity MCMC simulation (the chain number is 5 and the length of each chain is 4,000), the adaptive $\hat{f}(\mathbf{m})$-MCMC simulation using Strategy A (GP, $N_{ini}$=40, $N_a$=10, and $I_{max}$=10), and the adaptive $\hat{f}(\mathbf{m})$-MCMC simulation using Strategy B (PCE + GP, $N_{ini}$=40, $N_a$=10, and $I_{max}$=10). In this case, as the distribution of measurement error is not available, we use the following form of log-likelihood function (Vrugt, 2016) in the MCMC

simulations:

$$\mathcal{L}_{\log}(\mathbf{m}|\tilde{\mathbf{y}}) = -\frac{N_y}{2}\log\left\{\sum_{i=1}^{N_y}[f_i(\mathbf{m}) - \tilde{y}_i]^2\right\}, \tag{29}$$

where $f_i(\mathbf{m})$ and $\tilde{y}_i$ are the *i*th elements of model responses and actual measurements, respectively. In the adaptive $\hat{f}(\mathbf{m})$-MCMC simulation using Strategy A (GP), to account for the surrogate approximation uncertainty, we add random realizations of surrogate approximation errors to surrogate mean predictions to serve as model responses in the log-likelihood function of equation (29). In Figure 13, we compare the marginal posterior distributions of the unknown model parameters estimated by the three MCMC simulations. It is clear that with less than one percent of the number of high-fidelity model evaluations, the two proposed methods can obtain almost identical results as the high-fidelity MCMC simulation. Different from the results presented in Example 1, where the approach using Strategy A (GP) performs slightly worse than the approach using Strategy B (PCE + GP), here the two strategies exhibit similar performances. This is not surprising due to the rather low dimensionality of the target distribution.

[Figure 13 NEAR HERE]

As this case considers a real-world application, the true values of model parameters are not available. To verify the performances, we draw 50 posterior parameter samples obtained by the high-fidelity MCMC simulation and the adaptive $\hat{f}(\mathbf{m})$-MCMC simulation using Strategy B (PCE + GP), and depict the corresponding high-fidelity model responses (green curves) against the actual observational data (blue circles) in Figure 14. It is clear that both approaches (Figures 14a and 14b) can match the observations quite well.

[Figure 14 NEAR HERE]

## 4. Conclusions and Discussions

In this paper, we propose an adaptive approach that explicitly quantifies and gradually reduces the surrogate approximation error in surrogate-based MCMC simulations. Here, two strategies are proposed to quantify the surrogate approximation error. In the first strategy (Strategy A), we try to account for the surrogate approximation uncertainty by quantifying the uncertainty of the surrogate hyperparameters (or coefficients) with a Bayesian method. In this way, the high-fidelity model outputs are likely to be covered by the uncertainty ranges of the surrogate outputs. In the second strategy (Strategy B), a secondary surrogate is used to directly simulate the difference between the high-fidelity and the primary surrogate model outputs. Different from Strategy A, Strategy B generates a directional leap from the primary surrogate outputs to the corresponding high-fidelity outputs. As our concern is the posterior distribution, we employ an active learning process that gradually refines the surrogate in the high posterior density region. With enough iterations, the surrogate approximation error can be reduced to a small level.

To demonstrate the performance of the proposed method, we test three case studies involving high-dimensionality, multi-modality and a real-world application, respectively. It is shown that if the surrogate approximation error is not quantitatively considered in the MCMC simulations, the inversion results will be biased. Without increasing the number of high-fidelity model evaluations, Strategy A and Strategy B can both reduce the bias introduced by the surrogate approximation error. By comparing different adaptive surrogate-based MCMC simulations with or without considering surrogate approximation errors, we find that Strategy B that integrates two methods (i.e., PCE and GP) that complement each other has the best performance.

Here, we adopt MCMC to infer the unknown model parameters, the two strategies proposed in this work can also be applied with other inverse algorithms, e.g., ensemble Kalman filter and its variants. Moreover, although the PCE and GP surrogates are utilized and discussed in this work, the proposed framework can be also combined with other surrogate modeling techniques, e.g., deep learning, for better performances in specific problems (Mo et al., 2018; Zha et al., 2018). Nevertheless, there are some issues that have not been addressed in the present work. For example, in practice we should diagnose the structural inadequacy of the hydrological model (Gupta et al., 2008). Using multiple summary metrics that measure relevant parts of system behavior, we can gain information about how and where the model may be improved (Sadegh et al., 2015; Vrugt & Sadegh, 2013). In future works, we can simultaneously consider the surrogate and the model structural errors by combining the approach proposed in this work with methods that address model structural inadequacy, e.g., (Duan et al., 2007; Madadgar & Moradkhani, 2014; Xu & Valocchi, 2015; Xu et al., 2017; Ye et al., 2010; Zeng et al., 2018; Zeng et al., 2016; Zhang et al., 2019).

**Acknowledgments**

Computer codes and data used in this paper are available at:

https://www.researchgate.net/publication/329305979_Adaptive_Surrogate-Based_MCMC_Considering_Surrogate_Approximation_Error

This work is supported by the National Natural Science Foundation of China (grants 41807006, 41771254 and 41877465) and China Postdoctoral Science Foundation funded project (grant 2018M630680).

The authors would like to thank the editor and anonymous reviewers for their constructive comments and suggestions, which significantly improve the quality of this work. The authors would also like to thank Bruno Sudret from ETH Zurich for providing the UQLab toolbox, Maziar Raissi from Brown University for providing the MATLAB codes of GP, Jasper Vrugt from University of California, Irvine for providing the MATLAB codes of DREAM$_{(ZS)}$, respectively.

# Tables

**Table 1**. Prior ranges and true values of the contaminant source parameters in Example 1

| Parameter | Range | True value |
|---|---|---|
| $x_s [L]$ | [3-5] | 4.033 |
| $y_s [L]$ | [4-6] | 5.405 |
| $s_1 [MT^{-1}]$ | [0-8] | 1.229 |
| $s_2 [MT^{-1}]$ | [0-8] | 7.628 |
| $s_3 [MT^{-1}]$ | [0-8] | 4.327 |
| $s_4 [MT^{-1}]$ | [0-8] | 5.438 |
| $s_5 [MT^{-1}]$ | [0-8] | 0.293 |
| $s_6 [MT^{-1}]$ | [0-8] | 6.474 |

**Table 2**. Brief descriptions of different adaptive surrogate-based MCMC approaches

| Name | Brief description |
|---|---|
| PCE without error estimation | MCMC simulation using a PCE surrogate. Here the surrogate approximation error is not considered. |
| GP without error estimation | MCMC simulation using a GP surrogate. Here the surrogate approximation error is not considered. |
| Strategy A: PCE | MCMC simulation using a PCE surrogate. Here the surrogate approximation error is considered with Strategy A. |
| Strategy A: GP | MCMC simulation using a GP surrogate. Here the surrogate approximation error is considered with Strategy A. |
| Strategy A: GP + ASMO-PODE | MCMC simulation using a GP surrogate. Here the surrogate approximation error is considered with Strategy A, and the new design points are generated with the method proposed by Gong & Duan (2017). |
| Strategy B: PCE + GP | MCMC simulation using a PCE surrogate. Here we use a GP surrogate (only the mean prediction) to simulate and correct the approximation error of the primary PCE surrogate. |
| Strategy B: PCE + PCE | MCMC simulation using a PCE surrogate. Here we use another PCE surrogate to simulate and correct the approximation error of the primary PCE surrogate. |
| Strategy B: GP + PCE | MCMC simulation using a GP surrogate (only the mean prediction). Here we use a PCE surrogate to simulate and correct the approximation error of the primary GP surrogate. |
| Strategy B: GP + GP | MCMC simulation using a GP surrogate (only the mean prediction). Here we use another GP surrogate (only the mean prediction) to simulate and correct the approximation error of the primary GP surrogate. |

**Table 3.** Prior ranges and true values of the contaminant source parameters in Example 2

| Parameter | $x_s[L]$ | $y_s[L]$ | $S_s[MT^{-1}]$ | $t_{on}[T]$ | $t_{off}[T]$ |
|---|---|---|---|---|---|
| Range | [3-5] | [3-7] | [10-13] | [3-5] | [9-11] |
| True value | 3.854 | 5.999 | 11.044 | 4.897 | 9.075 |

**Figures**

**Figure 1.** Flow domain of the first and second examples. The point contaminant source (S) is located somewhere in the red dashed rectangle. The measurement location(s) for the first and second examples are represented by the blue circles and square, respectively.

**Figure 2.** Trace plots of the high-fidelity MCMC simulation (i.e., $f(\mathbf{m})$-MCMC, blue dots, in the first case study of Example 1), posterior mean estimates obtained from adaptive $\hat{f}(\mathbf{m})$-MCMC simulations using PCE without error estimation (red circles), Strategy A (PCE, red diamonds), and Strategy B (PCE + GP, red squares), respectively. Here the eight panels are for the eight contaminant source parameters, and the true values are represented by the black crosses. MCMC = Markov chain Monte Carlo, PCE = polynomial chaos expansion, and GP = Gaussian process.

**Figure 3.** Initial and added design points generated by the adaptive $\hat{f}(\mathbf{m})$-MCMC simulation using PCE without error estimation (in the first case study of Example 1). Here the eight panels are for the eight contaminant source parameters, and the true values are represented by the black crosses. MCMC = Markov chain Monte Carlo, and PCE = polynomial chaos expansion.

**Figure 4.** Initial and added design points generated by the adaptive $\hat{f}(\mathbf{m})$-MCMC simulation that uses Strategy A to account for surrogate approximation error of PCE (in the first case study of Example 1). Here the eight panels are for the eight contaminant source parameters, and the true values are represented by the black crosses. MCMC = Markov chain Monte Carlo, and PCE = polynomial chaos expansion.

**Figure 5.** Initial and added design points generated by the adaptive $\hat{f}(\mathbf{m})$-MCMC simulation that uses Strategy B (PCE + GP) to account for surrogate approximation error (in the first case study of Example 1). Here the eight panels are for the eight contaminant source parameters, and the true values are represented by the black crosses. MCMC = Markov chain Monte Carlo, PCE = polynomial chaos expansion, and GP = Gaussian process.

**Figure 6.** (a) The true $Y$ field generated from 20 KL terms (in the first case study of Example 1), (b) the posterior mean estimate of $Y$ field from the $f(\mathbf{m})$-MCMC simulation, and (c) the posterior mean estimate of $Y$ field from the adaptive $\hat{f}(\mathbf{m})$-MCMC simulation using Strategy B (PCE + GP). KL = Karhunen-Loève, MCMC = Markov chain Monte Carlo, PCE = polynomial chaos expansion, and GP = Gaussian process.

**Figure 7.** Evolution of the error of the surrogate at the true model parameters (in the first case study of Example 1) using the adaptive $\hat{f}(\mathbf{m})$-MCMC simulation with Strategy B (PCE + GP). Here the error index at the $y$ axis is defined in equation (21). MCMC = Markov chain Monte Carlo, PCE = polynomial chaos expansion, and GP = Gaussian process.

**Figure 8.** Initial and added design points generated by the adaptive $\hat{f}(\mathbf{m})$-MCMC simulation that uses Strategy A to account for surrogate approximation error of GP (in the first case study of Example 1). Here the eight panels are for the eight contaminant source parameters, and the true values are represented by the black crosses. MCMC = Markov chain Monte Carlo, and GP = Gaussian process.

**Figure 9.** Marginal PPDFs of the eight contaminant source parameters (in the first case study of Example 1) estimated by the $f(\mathbf{m})$-MCMC simulation (green curves), the adaptive $\hat{f}(\mathbf{m})$-

MCMC simulation using Strategy A (GP, red dashed curves), the adaptive $\hat{f}(\mathbf{m})$-MCMC simulation using ASMO-PODE and Strategy A (GP, magenta dash-dot curves), and the adaptive $\hat{f}(\mathbf{m})$-MCMC simulation using Strategy B (PCE + GP, blue dashed curves), respectively. Here the true values are represented by the black vertical lines. PPDF = posterior probability density function, MCMC = Markov chain Monte Carlo, GP = Gaussian process, and PCE = polynomial chaos expansion.

**Figure 10.** Trace plots of the $f(\mathbf{m})$-MCMC simulation (blue dots, in the second case study of Example 1) and posterior mean estimates obtained from the adaptive $\hat{f}(\mathbf{m})$-MCMC simulation using Strategy B (PCE + GP, red squares) for the eight contaminant source parameters. Here the true values are represented by the black crosses. MCMC = Markov chain Monte Carlo, PCE = polynomial chaos expansion, and GP = Gaussian process.

**Figure 11.** (a) The true $Y$ field (in the second case study of Example 1) generated from sequential Gaussian simulation without dimension reduction, (b) the posterior mean estimate of $Y$ field from the $f(\mathbf{m})$-MCMC simulation, and (c) the posterior mean estimate of $Y$ field from the adaptive $\hat{f}(\mathbf{m})$-MCMC simulation using Strategy B (PCE + GP). MCMC = Markov chain Monte Carlo, PCE = polynomial chaos expansion, and GP = Gaussian process.

**Figure 12.** Marginal PPDFs of the five contaminant source parameters (in Example 2) estimated by the $f(\mathbf{m})$-MCMC simulation (green curves) and the adaptive $\hat{f}(\mathbf{m})$-MCMC simulation using Strategy B (PCE + GP, blue dashed curves), respectively. Here the true values are represented by the black vertical lines. PPDF = posterior probability density function, MCMC = Markov chain Monte Carlo, PCE = polynomial chaos expansion, and GP = Gaussian process.

**Figure 13.** Marginal PPDFs of the five initial porosities (in Example 3) estimated by the $f(\mathbf{m})$-MCMC simulation (green curves), the adaptive $\hat{f}(\mathbf{m})$-MCMC simulation using Strategy A (GP, red dashed curves), and the adaptive $\hat{f}(\mathbf{m})$-MCMC simulation using Strategy B (PCE + GP, blue dashed curves), respectively. PPDF = posterior probability density function, MCMC = Markov chain Monte Carlo, PCE = polynomial chaos expansion, and GP = Gaussian process.

**Figure 14.** Posterior realizations (green curves) and posterior mean estimates (red dashed curves) of model predictions (in Example 3) obtained by (a) the $f(\mathbf{m})$-MCMC simulation and (b) the adaptive $\hat{f}(\mathbf{m})$-MCMC simulation using Strategy B (PCE + GP). The actual measurements are represented by the blue circles in each panel. MCMC = Markov chain Monte Carlo, PCE = polynomial chaos expansion, and GP = Gaussian process.

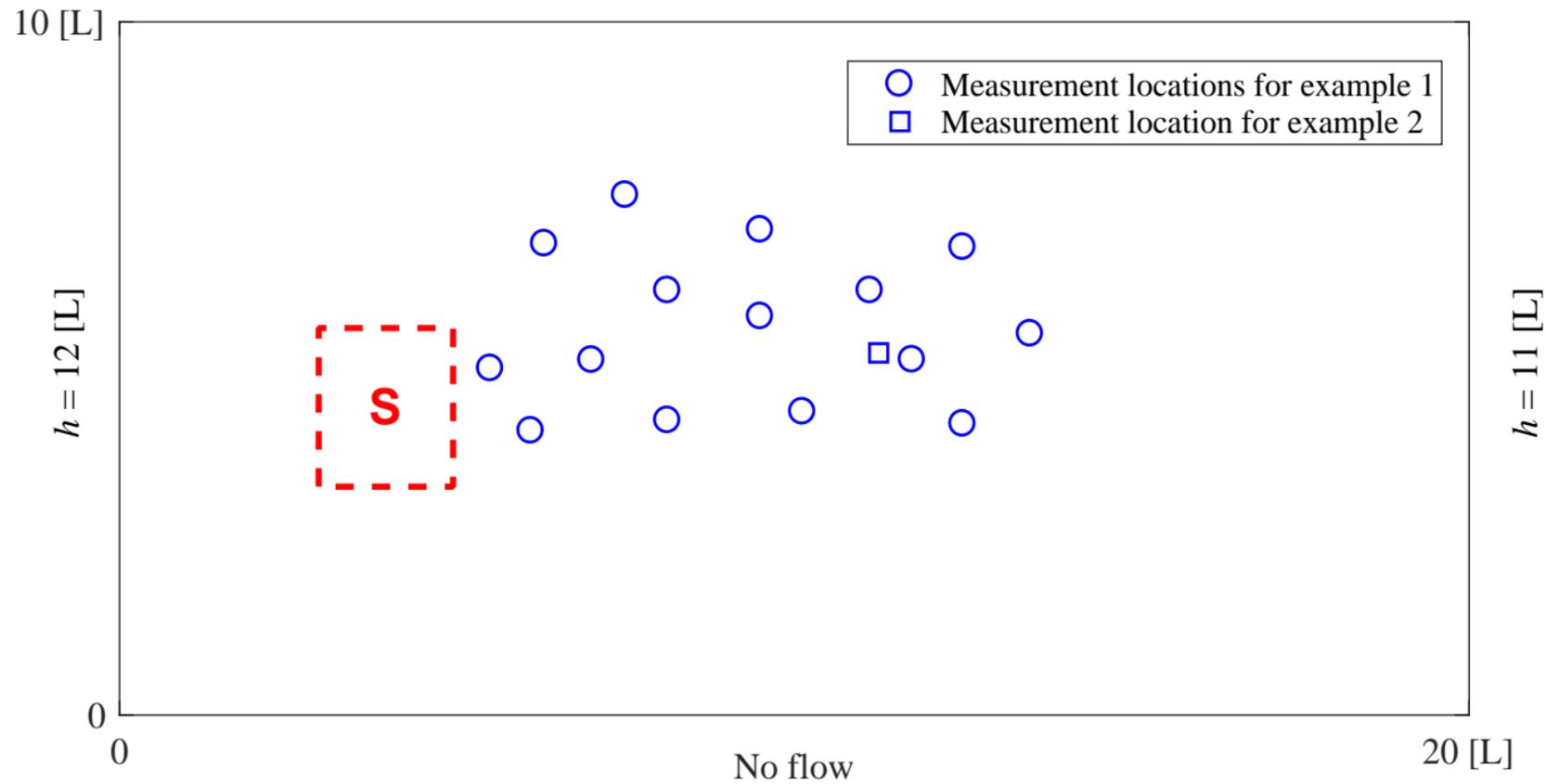

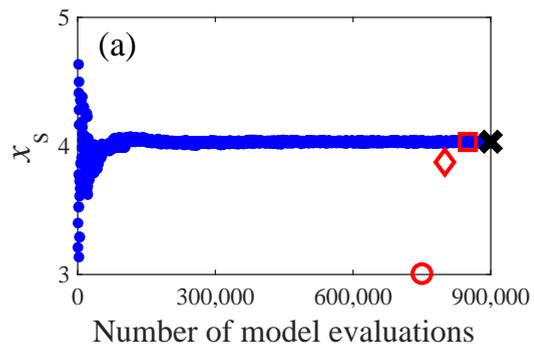 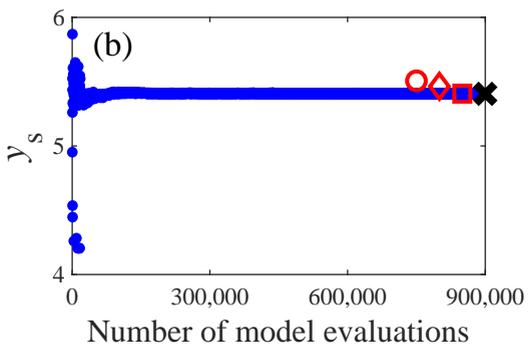 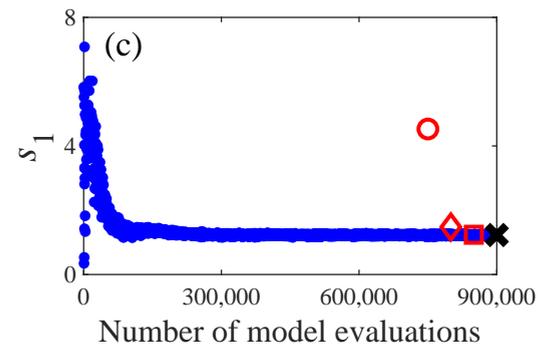
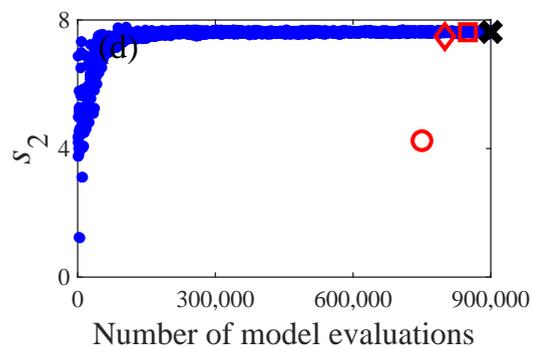 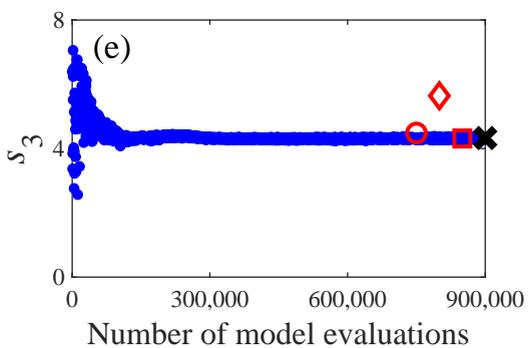 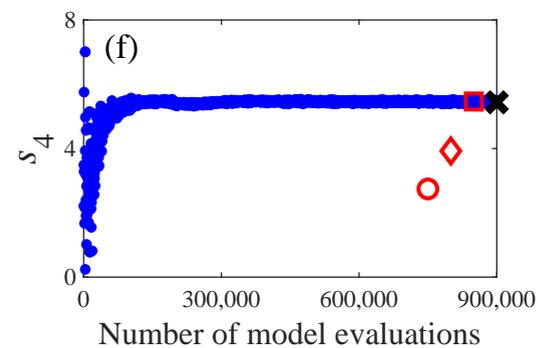
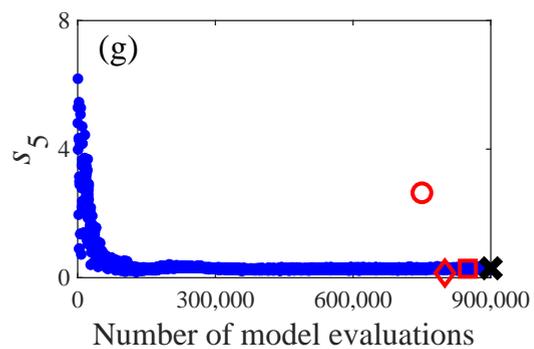 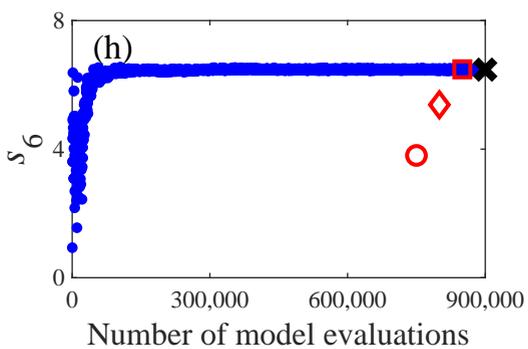 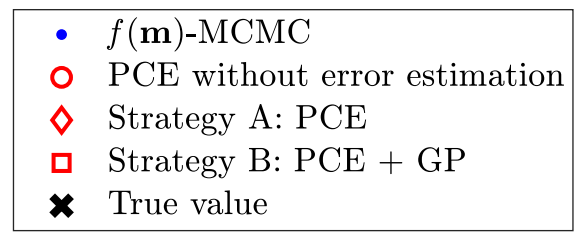

PCE without error estimation

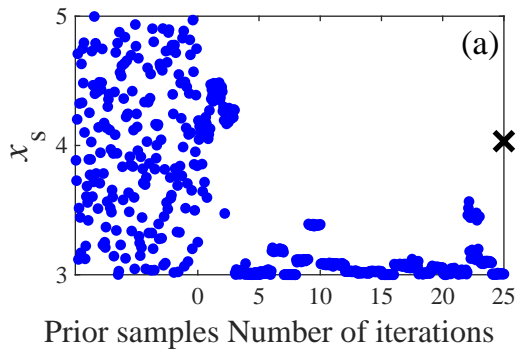
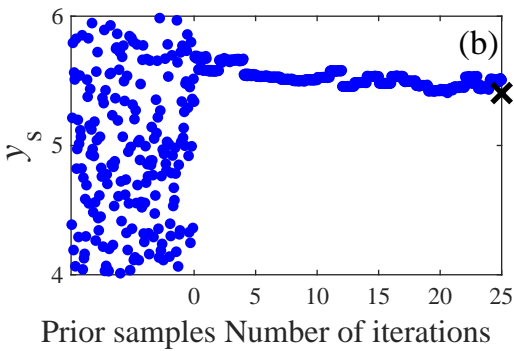
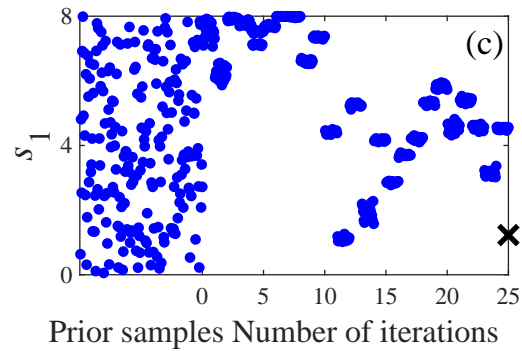
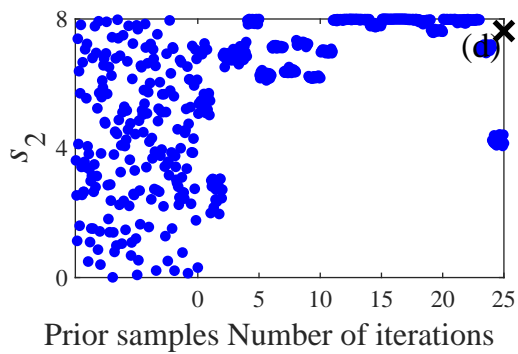
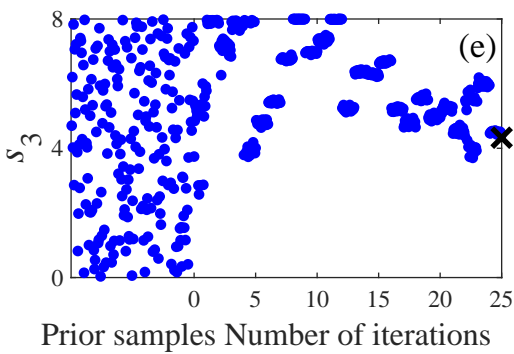
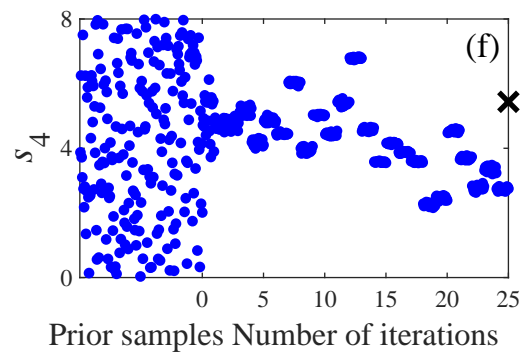
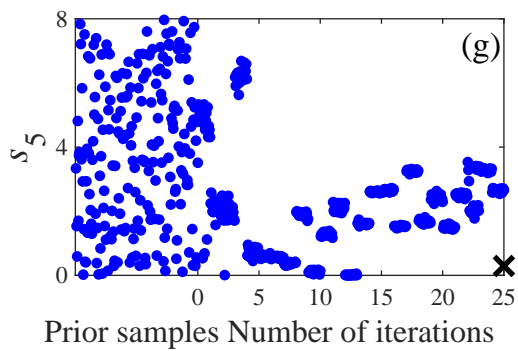
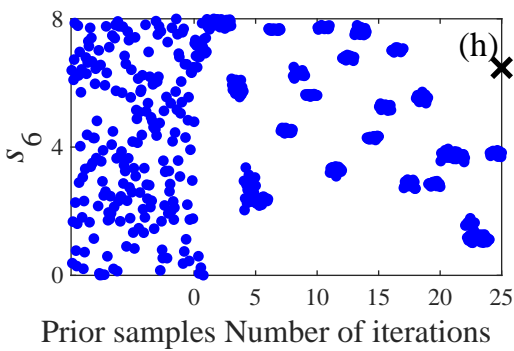

✕ True value

Number of $f(\mathbf{m})$ evaluations : 700

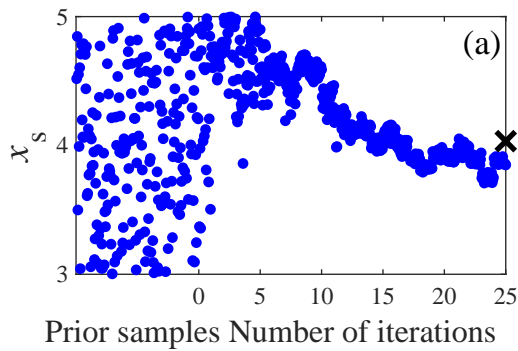
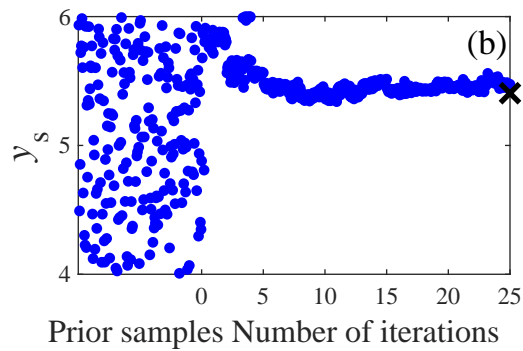
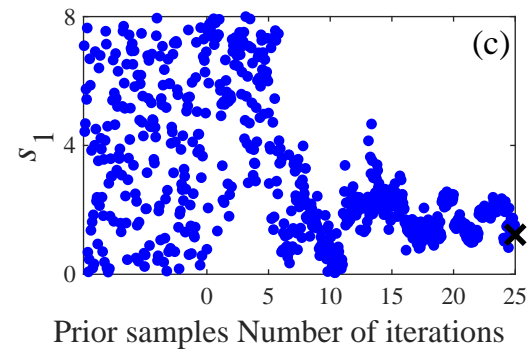
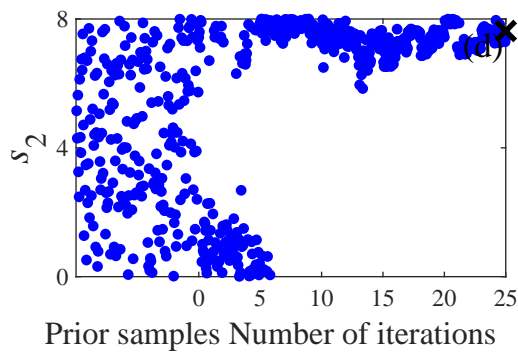
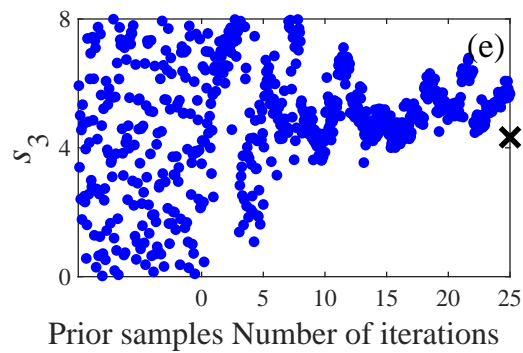
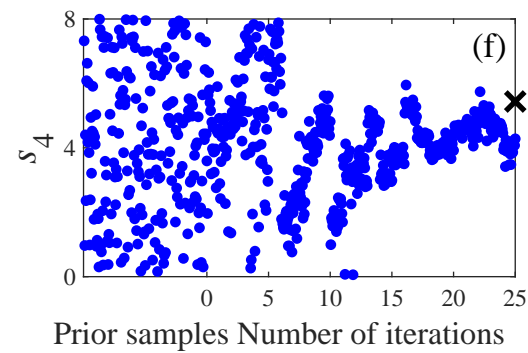
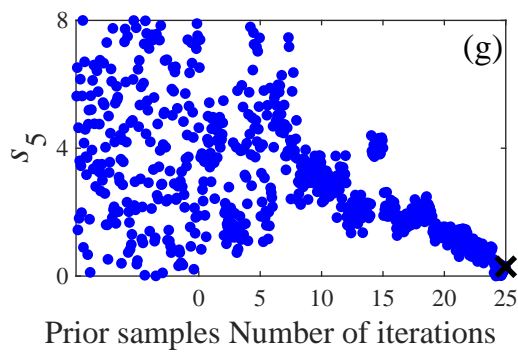
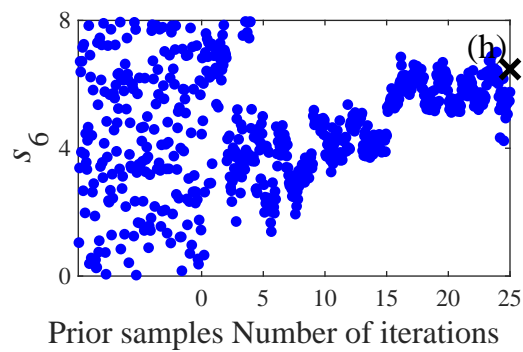

Strategy B: PCE + GP

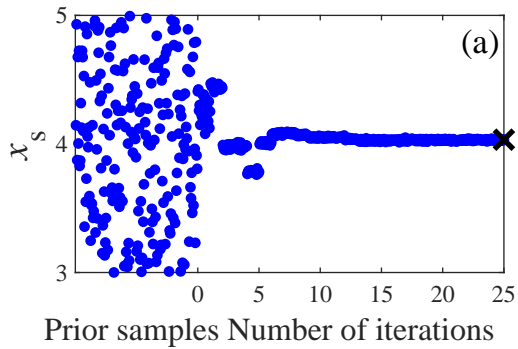
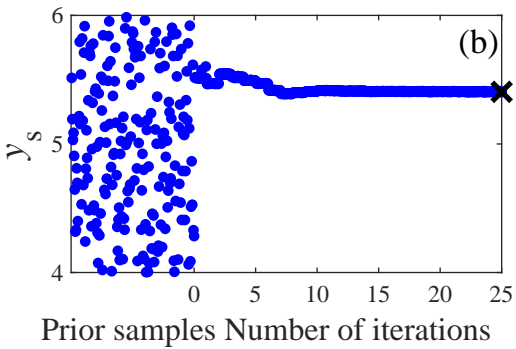
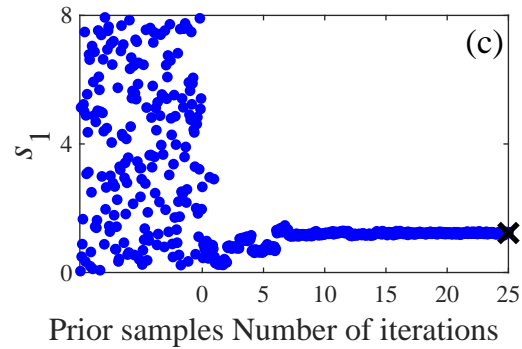
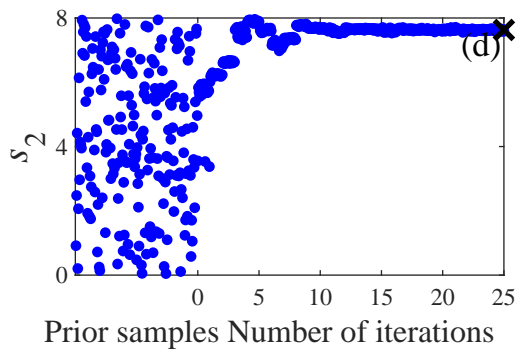
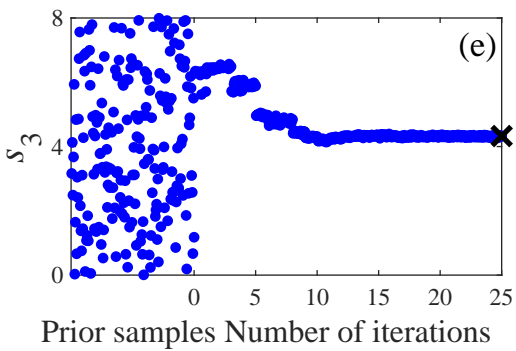
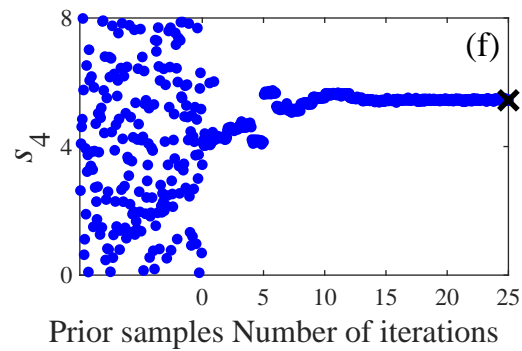
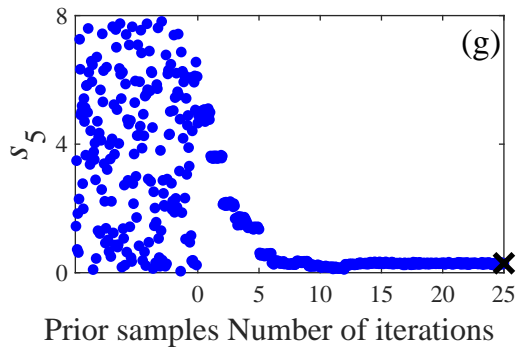
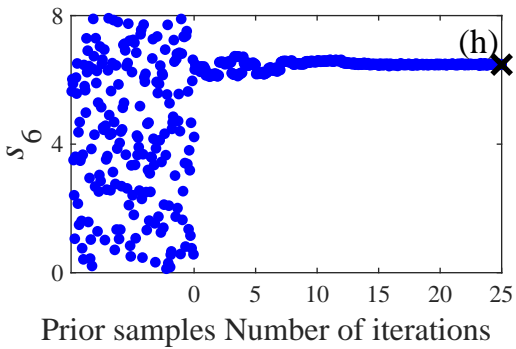

Number of $f(\mathbf{m})$ evaluations : 700

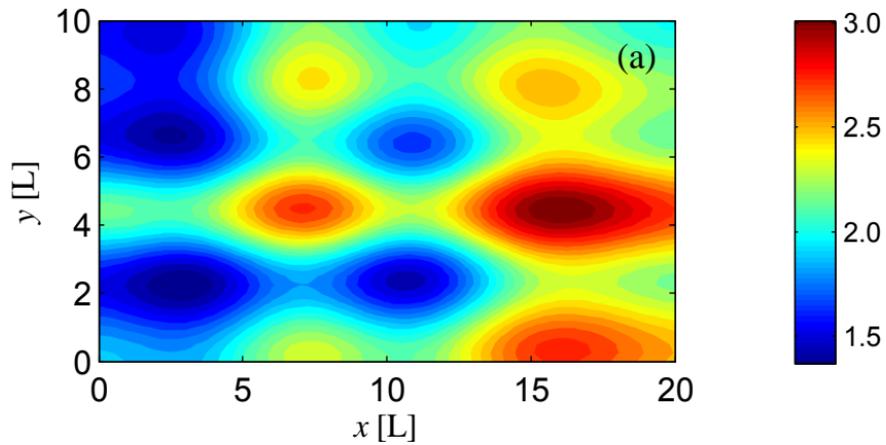
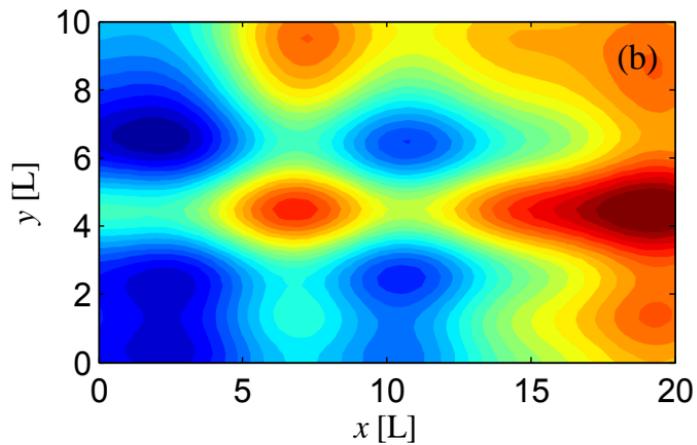
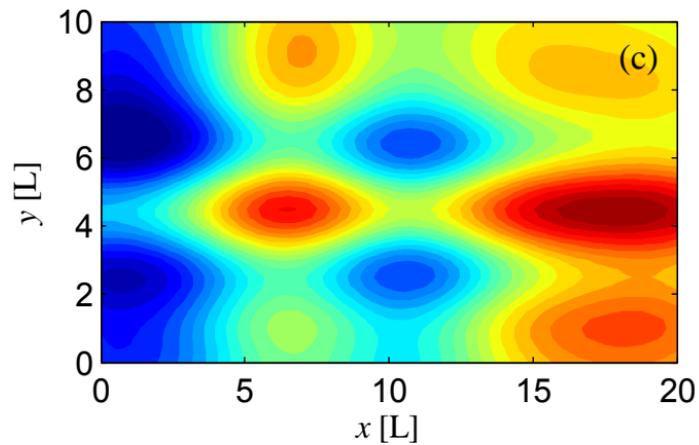

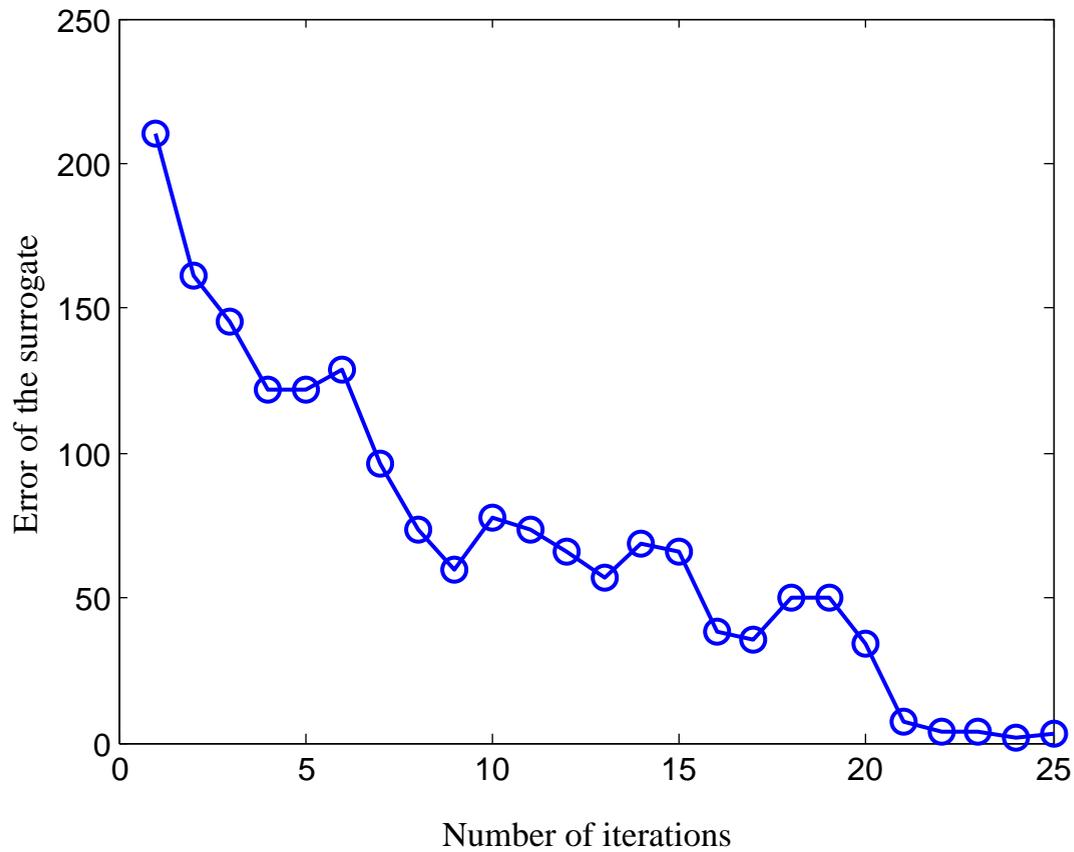

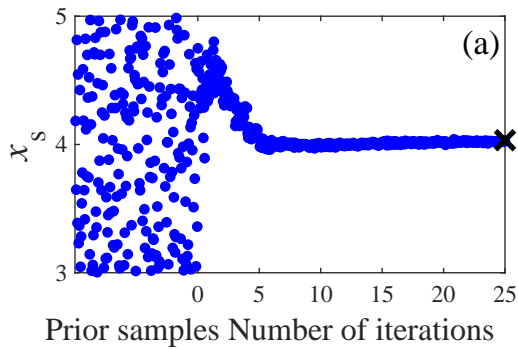 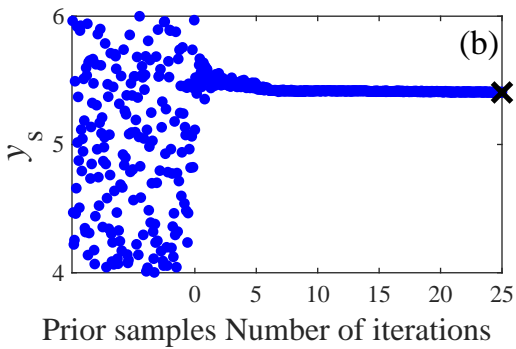 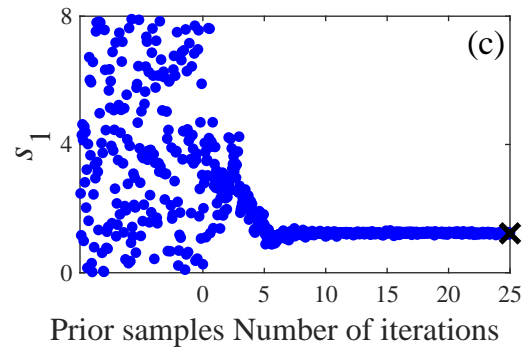
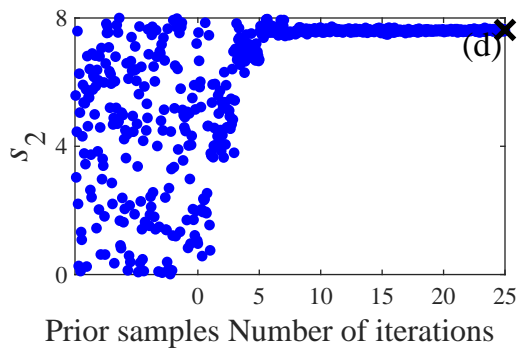 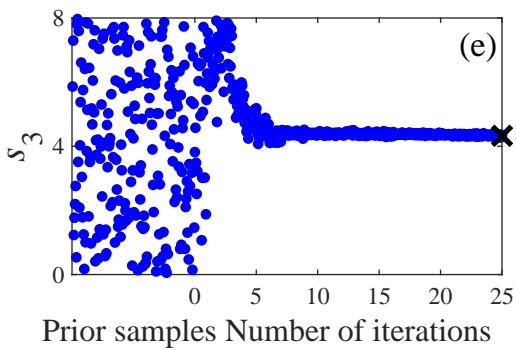 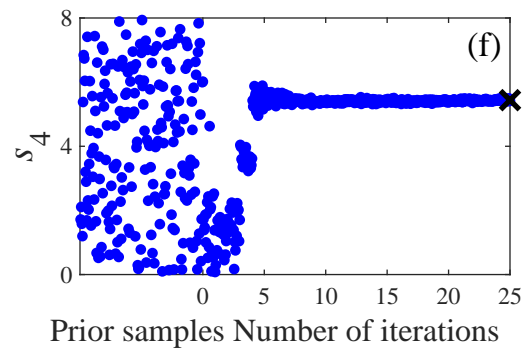
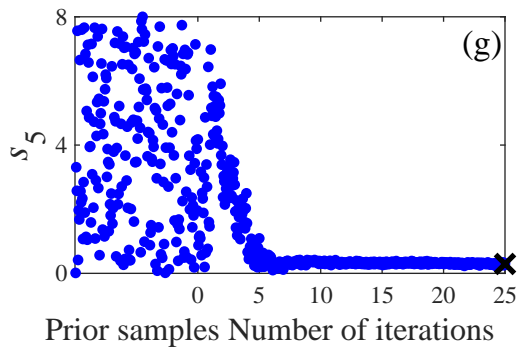 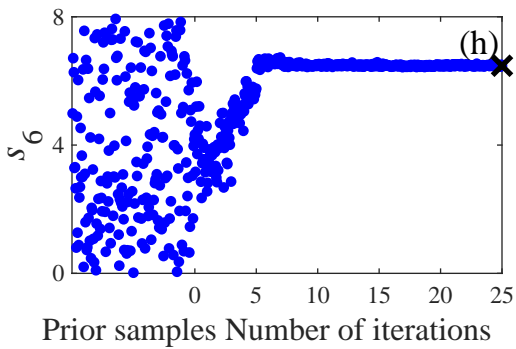

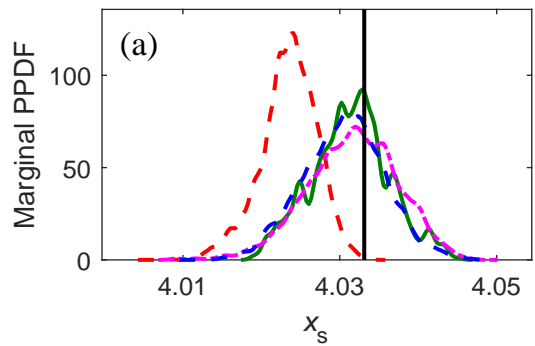 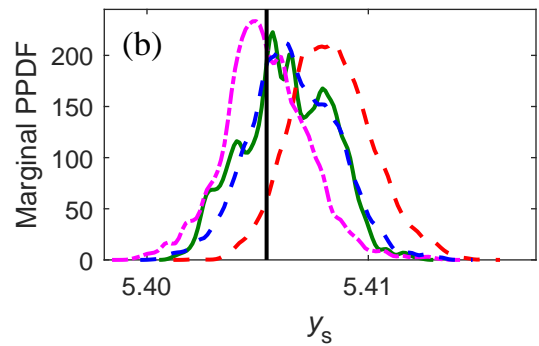 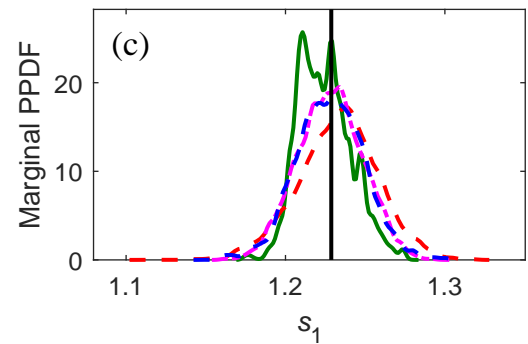
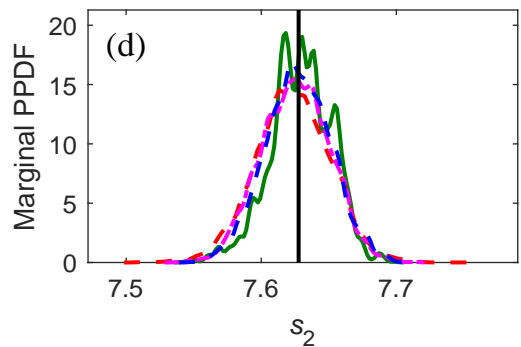 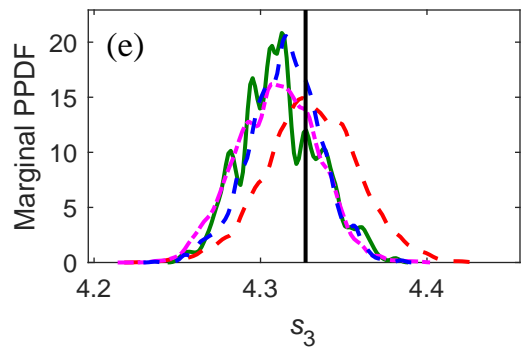 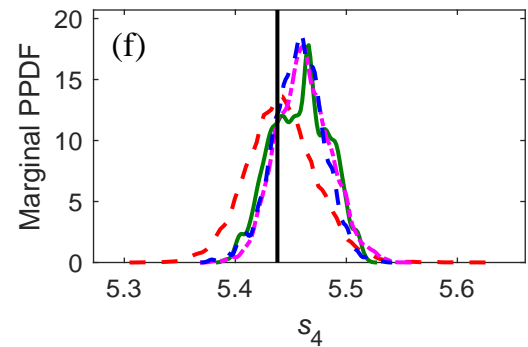
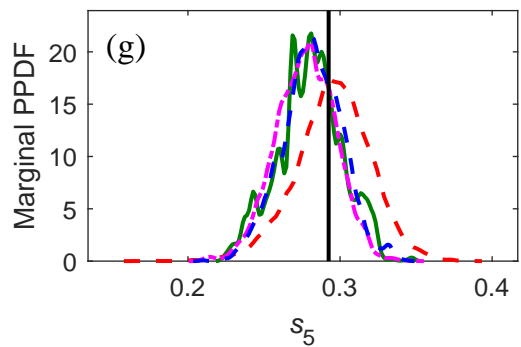 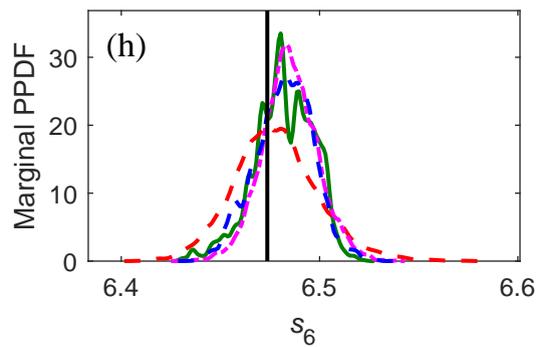 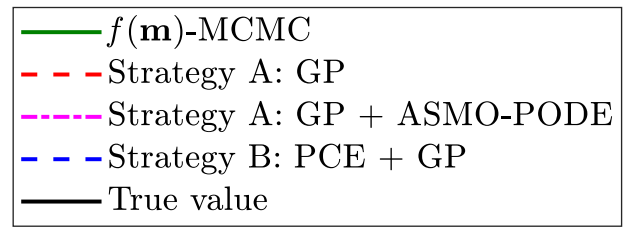

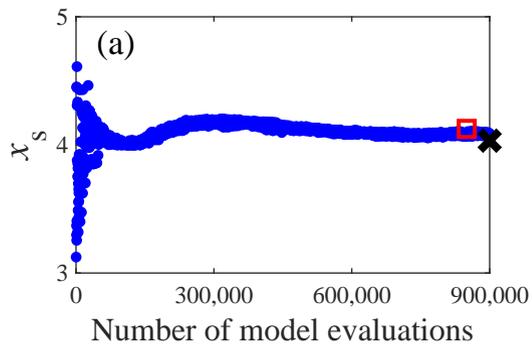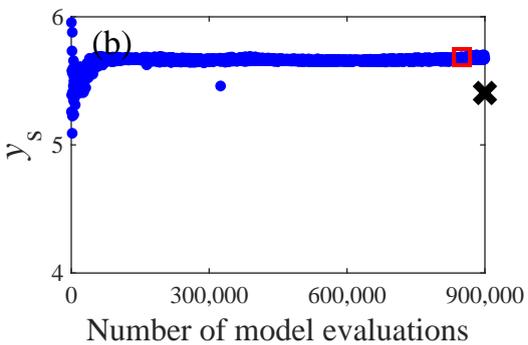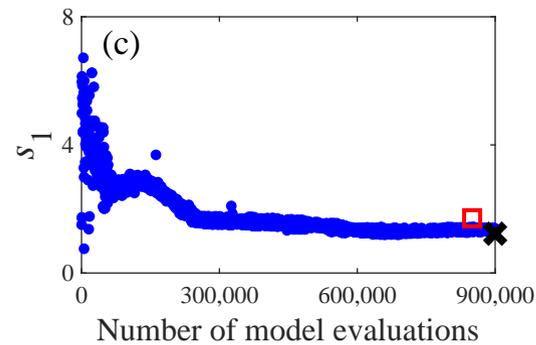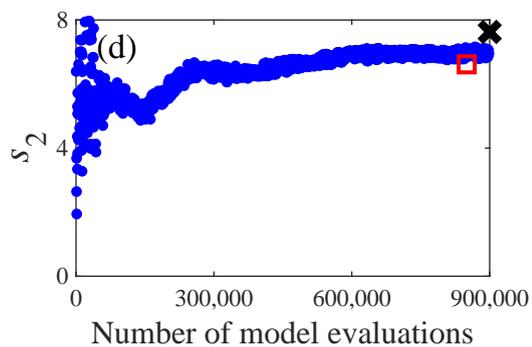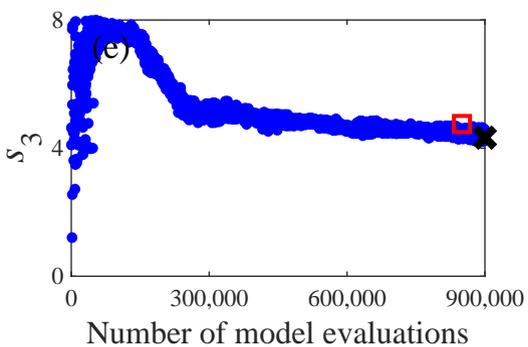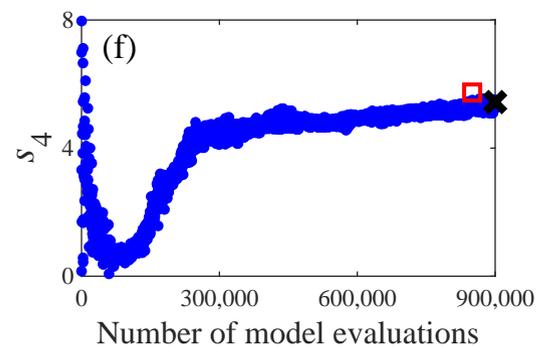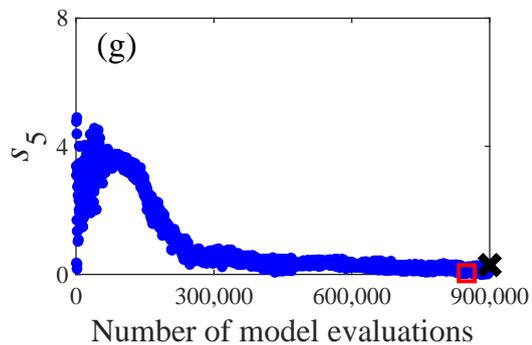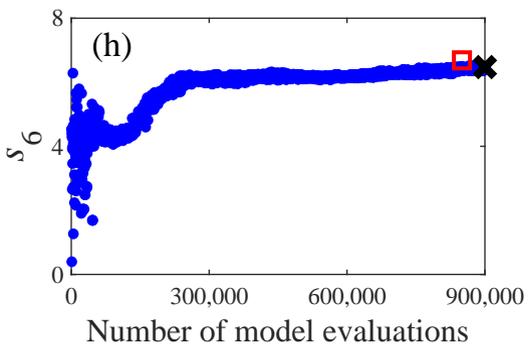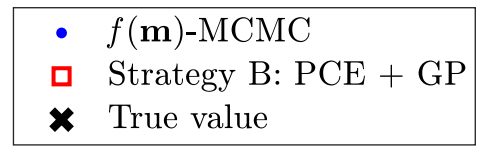

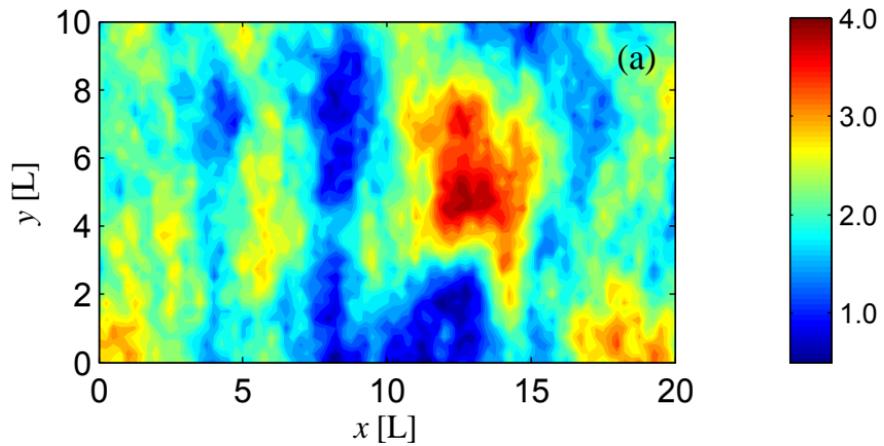
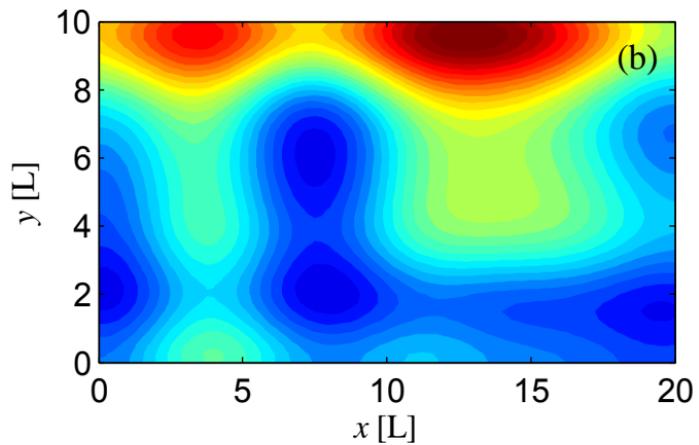
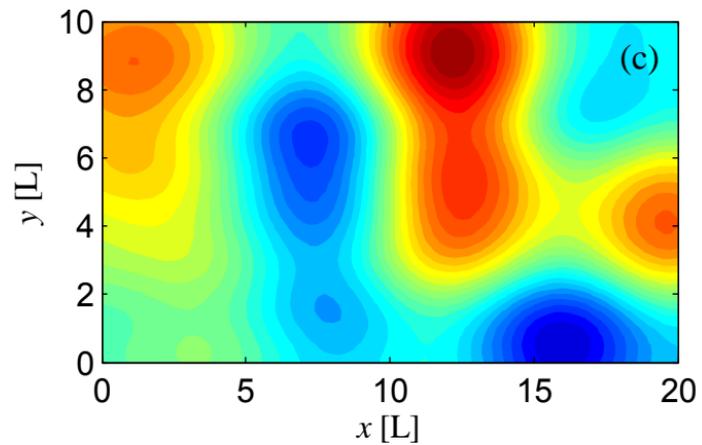

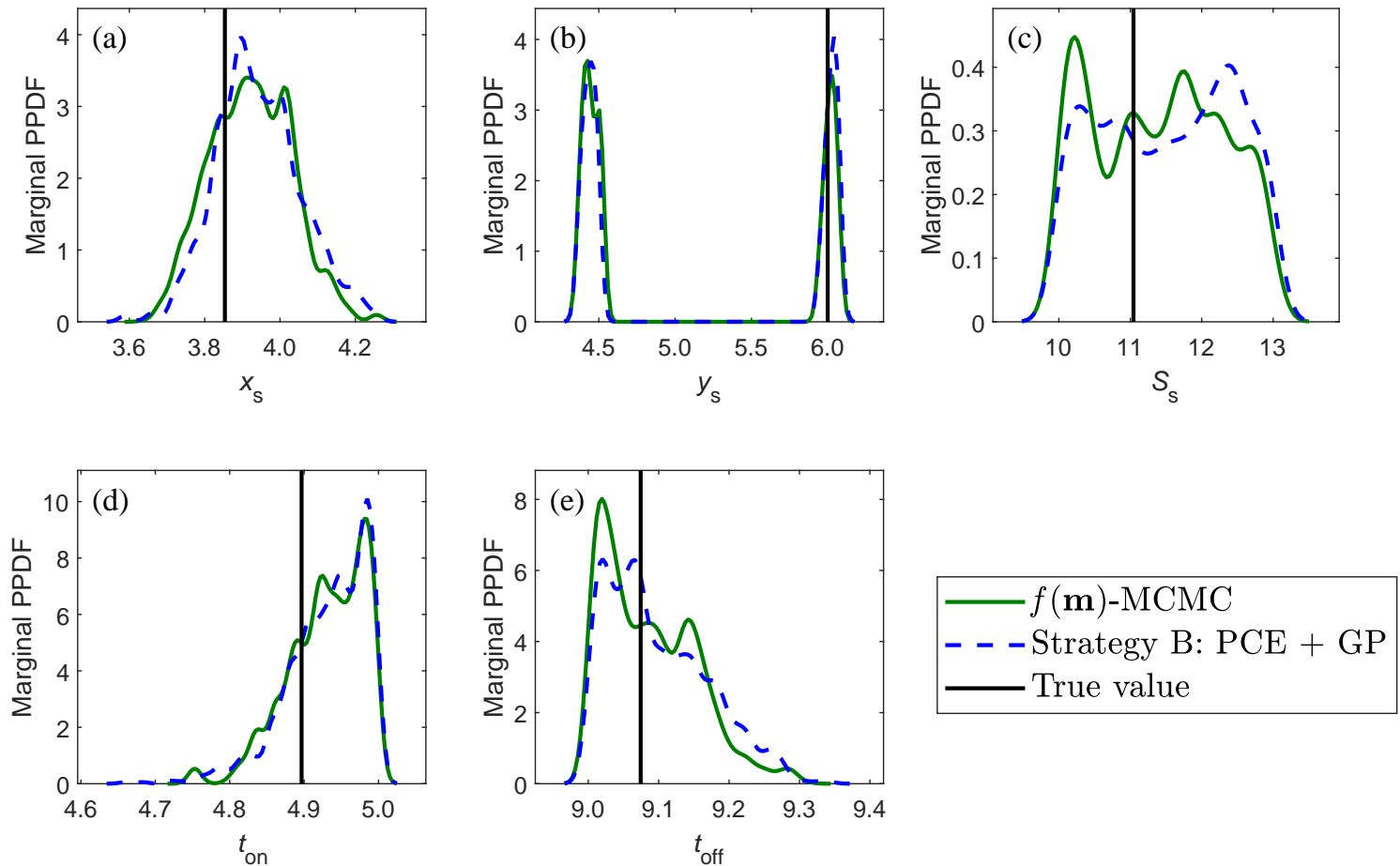

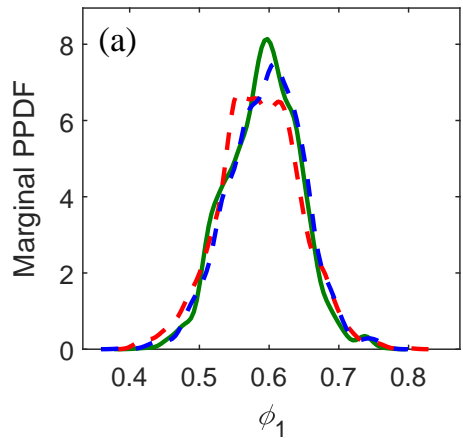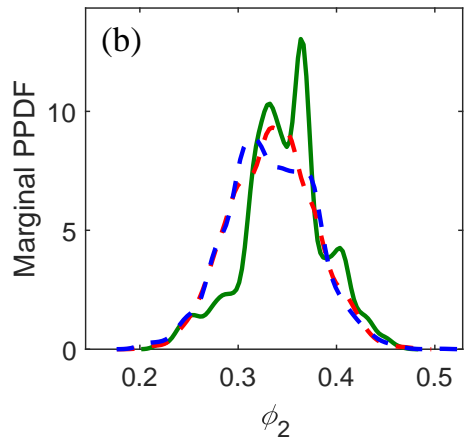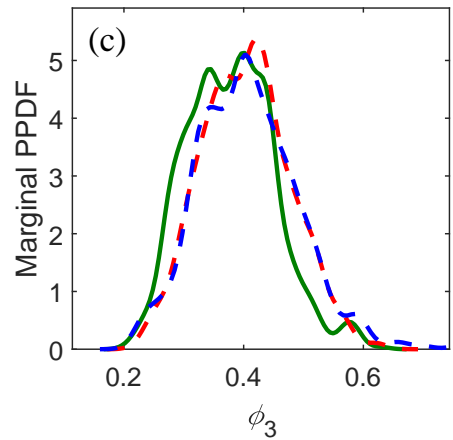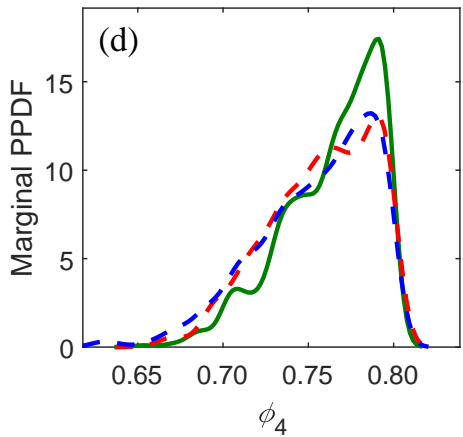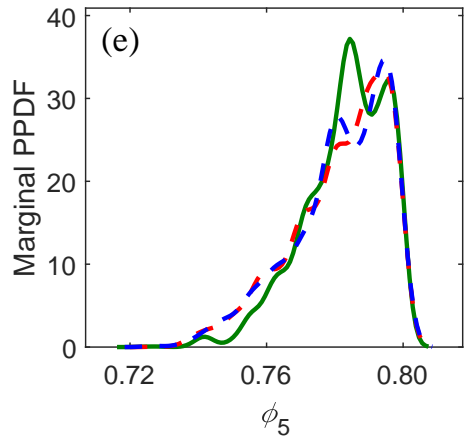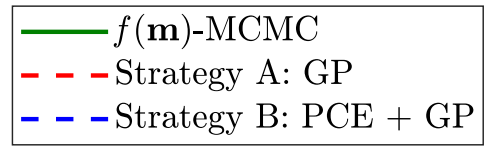

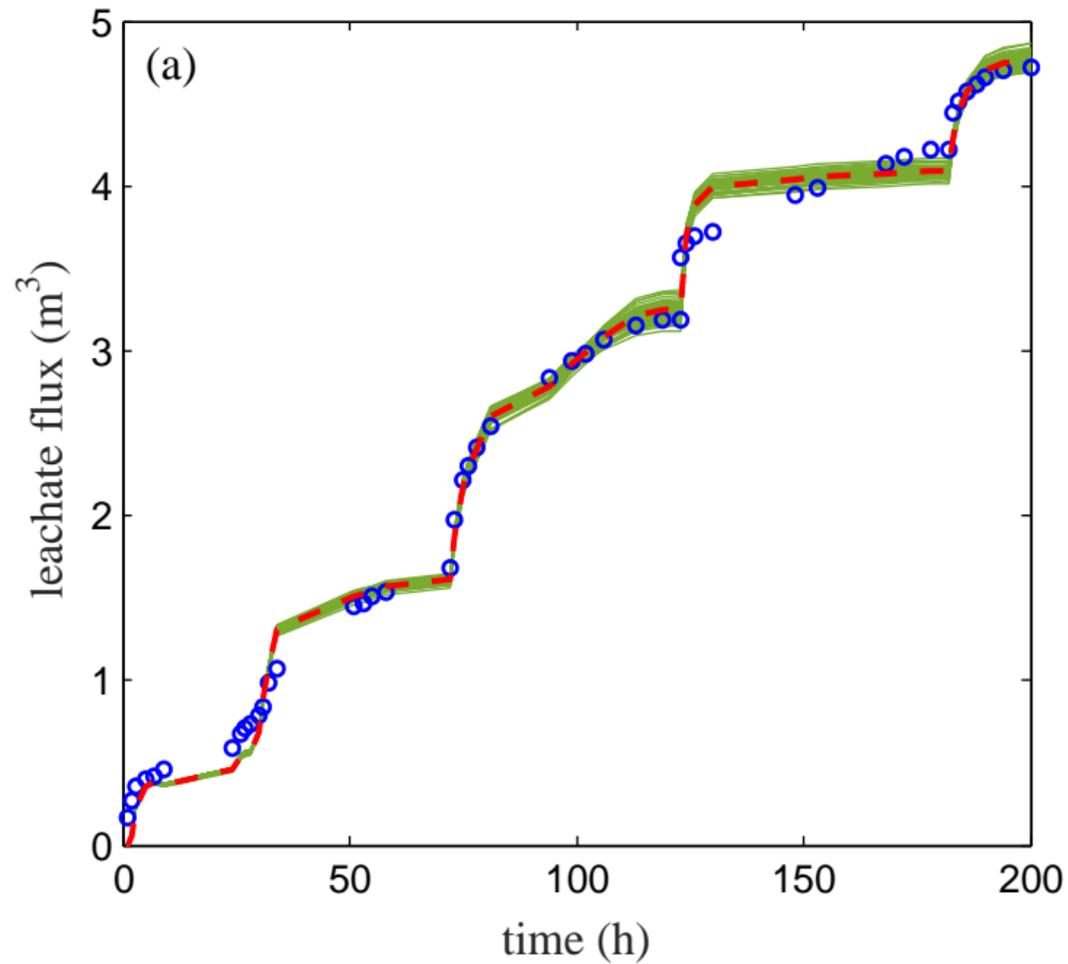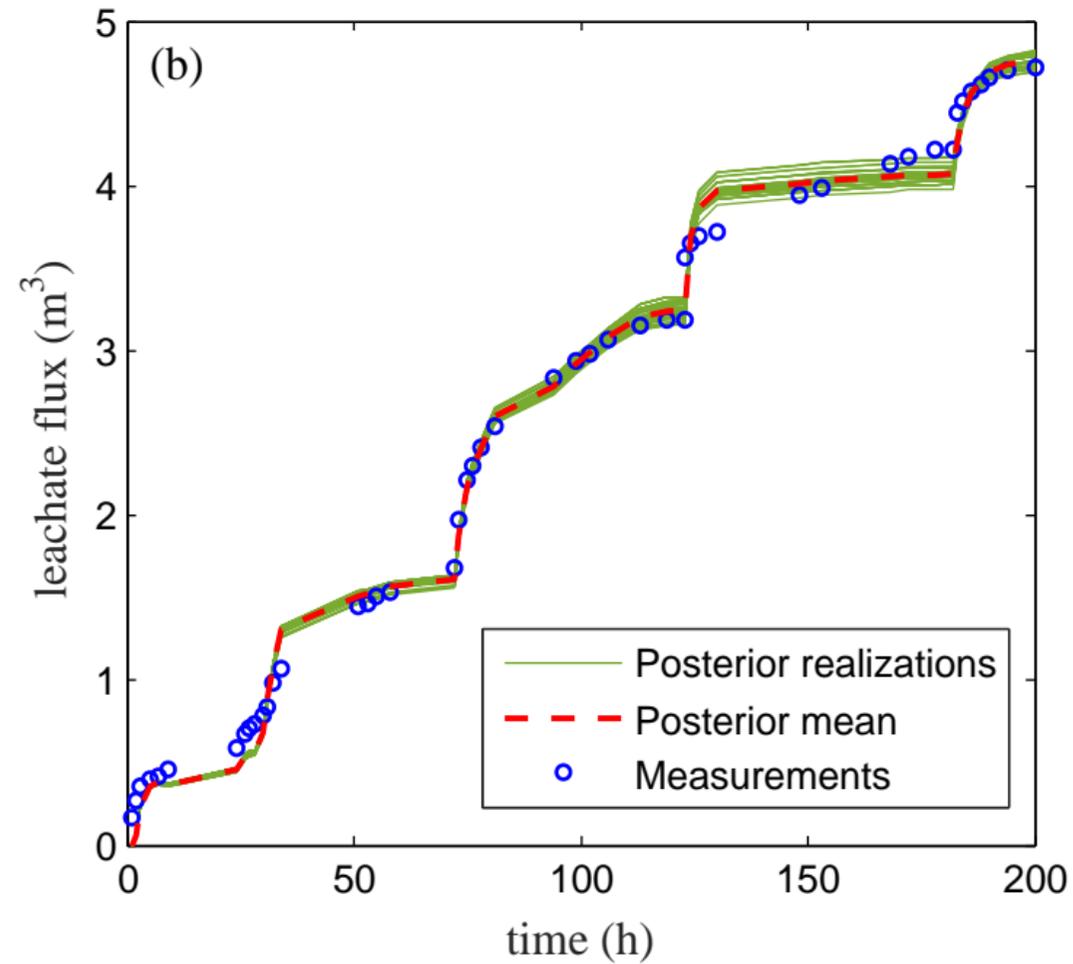